\documentclass[11pt]{article}

\usepackage{acl}

\usepackage{times}
\usepackage{latexsym}
\usepackage[T1]{fontenc}
\usepackage[utf8]{inputenc}
\usepackage{microtype}
\usepackage{inconsolata}

\usepackage{graphicx}
\usepackage{booktabs}
\usepackage{multirow}
\usepackage{amsmath}
\usepackage{tabularx}
\usepackage{placeins}
\usepackage{enumitem}
\usepackage[most]{tcolorbox}
\usepackage{xcolor}

\newtcolorbox{protocolbox}{
  enhanced jigsaw,
  breakable,
  colback=gray!3,
  colframe=gray!45,
  boxrule=0.35pt,
  arc=1.5pt,
  left=3pt,
  right=3pt,
  top=3pt,
  bottom=3pt,
  before skip=4pt,
  after skip=6pt
}

\newtcolorbox{annotationbox}{
  colback=gray!3,
  colframe=gray!55,
  boxrule=0.4pt,
  arc=2pt,
  left=5pt,
  right=5pt,
  top=5pt,
  bottom=5pt,
  breakable
}

\definecolor{fatalred}{RGB}{180,35,35}
\definecolor{misleadorange}{RGB}{190,110,20}

\newcommand{\fatal}[1]{\textcolor{fatalred}{\textbf{#1}}}
\newcommand{\mislead}[1]{\textcolor{misleadorange}{#1}}

\title{
\textsc{PhysDox}: Benchmarking LLMs on Physical Feasibility Auditing of
Physiological Sensing Protocols
}

\newcommand{\equalcontrib}{\textsuperscript{\ensuremath{\dagger}}}
\newcommand{\corresponding}{\textsuperscript{*}}

\author{
\textbf{He Liu}\textsuperscript{1,4}\equalcontrib,
\textbf{Boyuan Gu}\textsuperscript{2}\equalcontrib\corresponding,
\textbf{Shuaiqi Cheng}\textsuperscript{2,3},
\\
\textbf{Haiyang Sun}\textsuperscript{2},
\textbf{Siyu You}\textsuperscript{2},
\textbf{Xuming Hu}\textsuperscript{3}\corresponding
\\[4pt]
\textsuperscript{1}Donghua University, Shanghai, China \\
\textsuperscript{2}University of Electronic Science and Technology of China, Chengdu, China \\
\textsuperscript{3}The Hong Kong University of Science and Technology (Guangzhou), Guangzhou, China \\
\textsuperscript{4}The University of Hong Kong, Hong Kong SAR, China
\\[3pt]
\textit{Current affiliation of Boyuan Gu: Tsinghua Shenzhen International Graduate School,}\\
\textit{Tsinghua University, Shenzhen, China.}
\\[3pt]
\equalcontrib\ Equal contribution.
\qquad
\corresponding\ Corresponding authors.
}

\begin{document}

\maketitle

\begin{abstract}
Large language models (LLMs) increasingly assist in experimental design, yet fluent protocols often remain physically infeasible.
We introduce \textsc{PhysDox}, a physical feasibility auditing benchmark for biomedical protocols comprising a 683-sample expert-curated Gold set and a 5,000-sample Silver set across six sensing domains.
We formulate the task as a two-stage evaluation: severity detection classifying protocols as \textit{valid}, \textit{minor}, or \textit{fatal}, followed by the constraint-level diagnosis of fatal violations.
Evaluating 6 LLMs across 4 inference strategies yields a peak Stage-1 macro-F1 of only 53.0.
Moreover, strong oracle diagnosis collapses during end-to-end evaluation because errors are coupled across the detection and diagnosis stages.
Error analysis reveals \emph{scaffold bias}, where models conflate procedural completeness with physical validity.
Consequently, the six most implicit constraints exhibit roughly twice the expert--model performance gap of the six most explicit constraints (49.6 vs. 24.2 percentage points), while implicitness remains strongly associated with model miss rates (Spearman $\rho{=}0.81$, $p{<}0.01$).
Trace analysis of false negatives exposes a 54\%--46\% split between attention and judgment failures, ultimately demonstrating that protocol auditing demands calibrated feasibility reasoning rather than factual recall or longer rationales.
\end{abstract}

\section{Introduction}
\label{sec:intro}

Large language models (LLMs) are increasingly used to support scientific writing, experimental planning, and technical methodology drafting~\cite{brown2020language, smith2024ten, boiko2023autonomous, bran2024augmenting, jablonka2024gpt}. 
However, as LLMs enter complex engineering and biomedical domains, a critical vulnerability emerges: \emph{can we trust a fluent experimental protocol to be physically executable in the real world?}

\begin{figure}[t]
    \centering
    \includegraphics[width=1\linewidth]{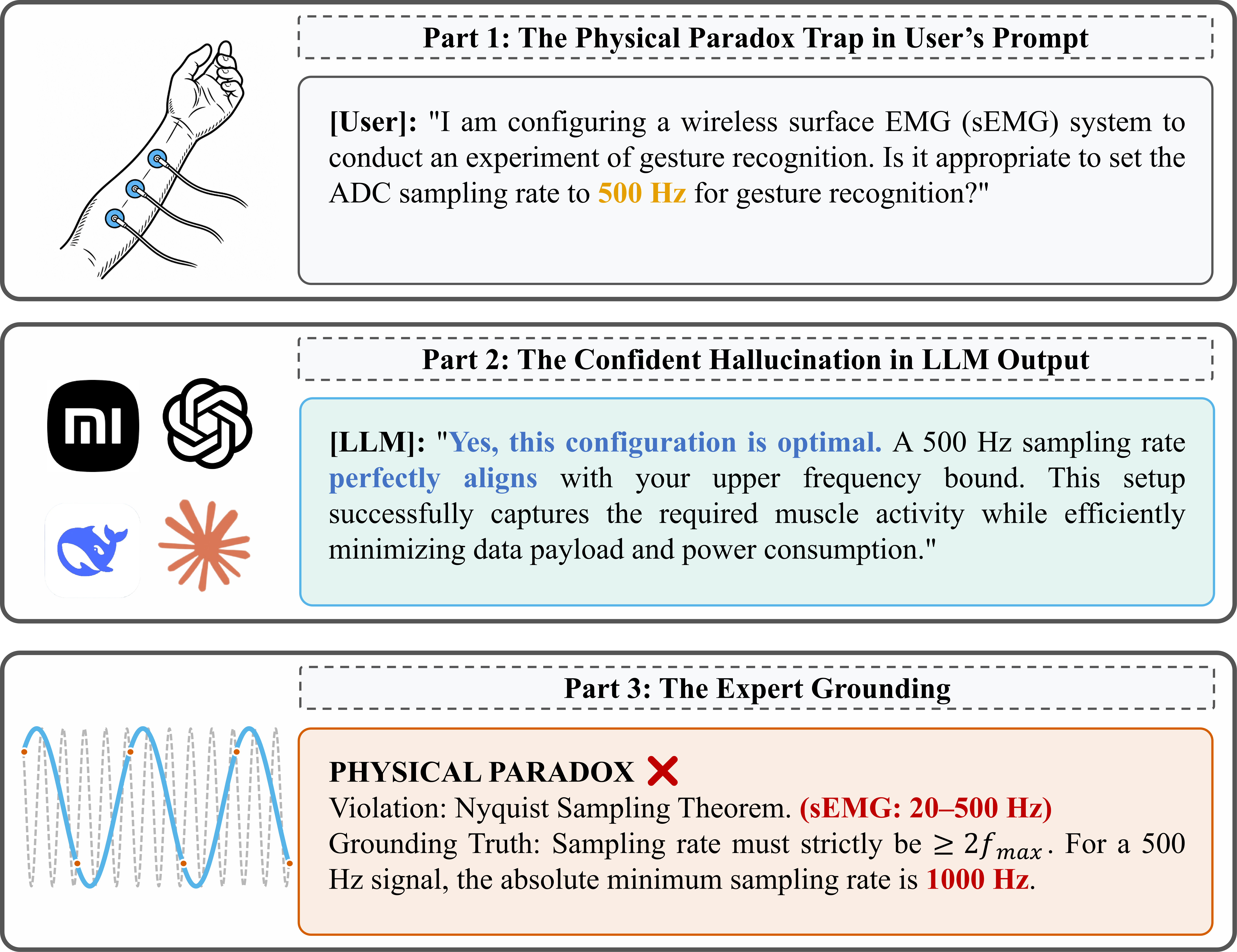}
    \caption{
    \textbf{The anatomy of a Physical Paradox.}
    The LLM confidently justifies a 500 Hz sampling rate for a signal containing 500 Hz sEMG components to ``optimize power'', while violating the Nyquist--Shannon sampling requirement.
    }\vspace{-1.0em}
    
    \label{fig:figure1}
\end{figure}

Consider a scenario where a researcher asks an LLM to configure a data acquisition pipeline for a surface electromyography (sEMG) signal with frequency components up to 500 Hz. 
As illustrated in Figure~\ref{fig:figure1}, the LLM confidently advises setting the sampling rate to 500 Hz to ``optimize power consumption.'' 
To a generalist, this may read as a reasonable engineering trade-off. 
To a signal-processing expert, it is a fatal violation of the Nyquist--Shannon sampling theorem, causing irreversible aliasing. 
We define this phenomenon as a \textbf{Physical Paradox}: a protocol that is semantically fluent, grammatically coherent, and procedurally complete, yet violates objective physical laws, hardware limits, or experimental feasibility constraints.

Recent efforts to evaluate LLM hallucinations have focused on \emph{semantic hallucinations}: fabricated facts, unsupported claims, or contradictory knowledge~\cite{lin2022truthfulqa, li2023halueval, min2023factscore, manakul2023selfcheckgpt, ravichander2025halogen}.
Meanwhile, physics and scientific reasoning benchmarks evaluate textbook-style problem solving~\cite{srivastava2023beyond, wang2024scibench, lu2024mathvista, feng2025physics}.
However, solving an isolated physics equation is fundamentally different from auditing a complete experimental protocol~\cite{liu2026bioprobench}: feasibility depends on whether sensing paths, sampling rates, hardware limits, and validation references are jointly executable.

Existing inference-time interventions---chain-of-thought prompting~\cite{wei2022chain}, self-consistency~\cite{wang2023selfconsistency}, and tool-augmented prompting~\cite{yao2023react, schick2023toolformer}---are commonly used to improve reasoning reliability.
However, physical feasibility auditing imposes a different requirement: a model must calibrate whether the complete protocol is \textit{valid}, contains a \textit{minor issue}, or is \textit{fatally infeasible}.
This makes protocol auditing a severity-calibration problem, not merely a reasoning-length problem.

To study this gap systematically, we introduce \textsc{PhysDox}, a benchmark for physical feasibility auditing of biomedical experimental protocols across physiological and non-contact sensing domains. 
Our core contributions are three-fold:
\begin{itemize}
\item \textbf{Benchmark.} We introduce \textsc{PhysDox}, a biomedical protocol auditing benchmark containing 5,000 Silver and 683 expert-curated Gold samples across six sensing domains.
\item \textbf{Task formulation.} We frame auditing as a two-stage evaluation: severity detection (\textit{valid}, \textit{minor}, \textit{fatal}) followed by fine-grained constraint diagnosis.
\item \textbf{Findings.} Evaluations expose \emph{scaffold bias}, where models conflate procedural completeness with physical validity; the six most implicit constraints show roughly twice the expert--model gap of the six most explicit constraints.
\end{itemize}

\section{The Physical Paradox Benchmark}
\label{sec:benchmark}

\begin{figure*}[t]
    \centering
    \includegraphics[width=1\linewidth]{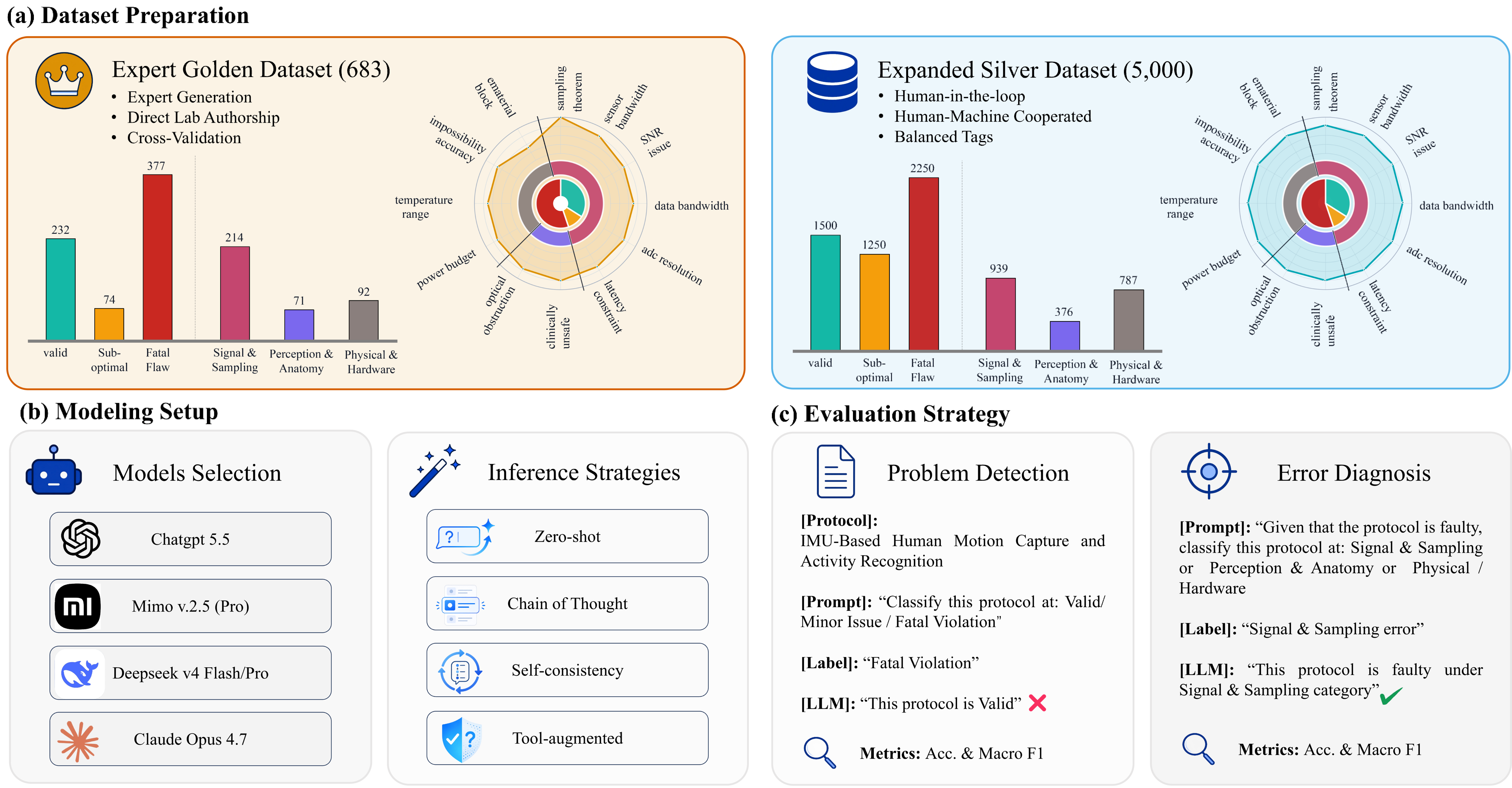}
    \caption{The architecture and profiling of \textsc{PhysDox}. The modeling panel illustrates representative model families; the full evaluation includes six variants from four families.}\vspace{-1em}
    \label{fig:pipeline}
\end{figure*}

To systematically probe LLMs' grounding in physical reality, we introduce the Physical Paradox Benchmark (\textsc{PhysDox}). As illustrated in our dual-track methodology (Figure~\ref{fig:pipeline}), the benchmark comprises a 5,000-case silver set generated via human-AI collaboration to evaluate breadth, and a meticulously curated 683-case gold set authored directly by domain experts. The Gold set is expert-curated to ensure physically grounded labels and stricter feasibility boundaries.

\subsection{Targeted Domains \& Tri-State Severity}
\label{sec:domains_severity}

To prevent LLMs from relying on superficial textual memorization, \textsc{PhysDox} focuses on six highly constrained physiological and non-contact sensing modalities: sEMG, FMCW Radar, EEG, IMU, PPG/ECG, and Throat Microphones. 

Crucially, rather than treating experimental design as a binary correct/incorrect task, we annotate protocols with a rigorous tri-state severity label to test fine-grained physical reasoning:
\begin{itemize}
    \item \textbf{Valid:} Protocols that perfectly adhere to both physical laws and engineering standards.
    \item \textbf{Minor Issue:} Configurations that are physically possible but exhibit poor engineering judgment.
    \item \textbf{Fatal Violation:} Protocols containing a core feasibility violation that renders the stated experiment non-executable under its specified physical, hardware, anatomical, clinical, safety, or validation conditions.
    
\end{itemize}

\subsection{Taxonomy of Fatal Violations}
\label{sec:taxonomy}

We further deconstruct the \textbf{Fatal Violations} into three orthogonal dimensions of hardware-grounded blind spots. This taxonomy aligns directly with our annotation framework, exposing exactly where LLMs' textual heuristics collapse against physical constraints:

\paragraph{1. Signal \& Sampling.}
Paradoxes violating discrete-time signal processing, information theory, and digital acquisition limits.
\textit{Sampling Theorem \& ADC Resolution:} Configuring a $<1000$ Hz ADC for a sEMG signal with a 500 Hz high-frequency component (triggering aliasing), or claiming to resolve micro-volt EEG signals using a coarse 8-bit ADC without an analog preamplifier.
\textit{Bandwidth, SNR \& Latency:} Contradictions involving sensor/data bandwidth bottlenecks, unfeasible Signal-to-Noise Ratios (SNR) for weak physiological signals, or latency inconsistencies in real-time pipelines.

\paragraph{2. Perception \& Anatomy.}
Severe spatial, kinematic, or line-of-sight contradictions in sensor deployment.
\textit{Anatomical Misplacements:} Placing sEMG electrodes on the biceps (upper arm) to classify fine finger flexions, entirely bypassing the relevant forearm musculature.
\textit{Clinically Unsafe:} Frequently utilizing invasive measurements (e.g., invasive arterial blood pressure) for healthy volunteers, causing severe safety concerns.

\paragraph{3. Physical / Hardware.}
Contradictions involving electromagnetic wave propagation, environmental limits, or thermodynamic constraints.
\textit{Material Block (EM Propagation):} Positioning a 77 GHz mmWave radar behind a 5 mm aluminum plate for ``shielding,'' ignoring that the skin depth of metal strictly prohibits RF penetration.
\textit{Impossible Accuracy \& Component Limits:} Claiming 0.01 mmHg accuracy for cuffless blood pressure without calibration, or proposing system configurations that mathematically violate objective power budgets and operational temperature ranges.
\vspace{-0.5em}
\subsection{Gold-set quality control}
The Gold set contains 683 expert-curated protocols with consensus labels over valid, minor issue, and fatal violation. Each protocol was independently reviewed by two annotators with biomedical-sensing and signal-processing background; disagreements were resolved by senior-expert adjudication. The two annotators achieved 91.4\% raw agreement on tri-state severity classification (Cohen's $\kappa=0.85$) and 93.6\% agreement on the binary fatal gate ($\kappa=0.87$). The Gold and Silver sets are fully disjoint: Gold is used as the authoritative benchmark for safety conclusions, while Silver supports large-scale trend analysis.

\vspace{-0.3em}

\section{Experiments and Analysis}
\label{sec:experiments}

\vspace{-0.5em}

\subsection{Experimental Setup}
\label{sec:setup}

Evaluations enforce JSON formatting, retrying malformed outputs to guarantee complete predictions for all runs. Zero-shot, CoT, and tool-augmented strategies use temperature 0.1. Self-consistency samples three temperature-0.7 outputs per item and aggregates predictions by vote count; three-way ties are resolved by the deterministic aggregation order implemented in our evaluation script. Tool augmentation appends a local deterministic rule-check report covering a subset of computable hardware and physical constraints. The report is intentionally treated as noisy supporting evidence rather than ground truth: several constraints, such as SNR floors, temperature limits, and latency constraints, require manual or contextual assessment and are not fully solved by the rule base.

\vspace{-0.3em}
\paragraph{Models.}
We evaluate six LLMs (GPT-5.5, Claude Opus 4.7, DeepSeek V4 Flash/Pro, MiMo v2.5/Pro) across four inference strategies: zero-shot, CoT, self-consistency, and tool augmentation. This matrix isolates both inherent model capabilities and the efficacy of standard reasoning enhancements on physical hallucination detection.
\vspace{-0.4em}
\paragraph{Datasets.}
Experiments use two \textsc{PhysDox} splits. The 5,000-sample Silver set serves solely as a diagnostic testbed for trend analysis; because it was partially generated via MiMo v2.5 Pro, we anchor all final benchmarking and safety conclusions on the 683-sample expert-curated Gold set. Gold protocols feature strict Stage-1 labels (\textit{valid}, \textit{minor}, \textit{fatal}), with fatal cases further annotated with coarse taxonomy and fine-grained constraints for Stage 2.
\vspace{-0.4em}
\paragraph{Metrics.}
Stage 1 evaluates accuracy, fatal-gate recall, and macro-F1, our primary metric to account for Gold-set class imbalance. Stage 2 reports oracle and end-to-end accuracy for both the 3-way taxonomy and 12-way constraint diagnosis. Main text tables focus on core Gold metrics; exhaustive per-class and strategy-level breakdowns are detailed in Appendix D.

\begin{table*}[t]
\centering
\scriptsize
\setlength{\tabcolsep}{3.2pt}
\renewcommand{\arraystretch}{1.06}
\resizebox{\linewidth}{!}{%
\begin{tabular}{ll ccc ccc cc ccc}
\toprule
\multirow{2}{*}{\textbf{Model}} & \multirow{2}{*}{\textbf{Strategy}} 
& \multicolumn{3}{c}{\textbf{Silver 5k S1}} 
& \multicolumn{3}{c}{\textbf{Gold 683 S1}} 
& \multicolumn{2}{c}{\textbf{Gold 3-way}} 
& \multicolumn{3}{c}{\textbf{Gold Fine-12}} \\
\cmidrule(lr){3-5}\cmidrule(lr){6-8}\cmidrule(lr){9-10}\cmidrule(lr){11-13}
& & \textbf{Acc.} & \textbf{M-F1} & \textbf{Fatal R.} 
& \textbf{Acc.} & \textbf{M-F1} & \textbf{Fatal R.} 
& \textbf{Oracle} & \textbf{E2E} 
& \textbf{Oracle} & \textbf{E2E} & \textbf{Drop}$\downarrow$ \\
\midrule
\multirow{4}{*}{GPT-5.5} 
& Zero-Shot & 54.3 & 43.2 & 82.7 & 53.9 & 37.8 & 83.0 & 50.7 & 43.2 & 62.9 & 52.5 & 10.3 \\
& CoT & 54.8 & 43.9 & 81.4 & 51.7 & 37.7 & 77.7 & 52.3 & 41.4 & 61.0 & 49.3 & 11.7 \\
& Self-Cons. & 54.3 & 43.1 & 82.8 & 54.6 & 38.5 & \textbf{83.8} & 53.1 & \textbf{45.4} & 65.8 & 56.2 & \textbf{9.5} \\
& Tool-Aug. & 54.6 & 43.7 & 81.5 & 50.2 & 36.9 & 74.8 & 56.0 & 43.5 & 65.0 & 51.7 & 13.3 \\
\cmidrule(lr){1-13}
\multirow{4}{*}{Claude Opus 4.7} 
& Zero-Shot & 64.6 & 57.9 & 87.3 & 56.7 & 51.7 & 65.0 & 58.1 & 42.4 & 66.0 & 46.7 & 19.4 \\
& CoT & 65.0 & 60.0 & 85.2 & 52.9 & 49.3 & 57.6 & 55.7 & 37.4 & 68.4 & 45.6 & 22.8 \\
& Self-Cons. & 64.1 & 57.5 & 86.8 & 57.2 & 51.4 & 67.6 & 58.9 & 44.0 & 69.2 & 49.9 & 19.4 \\
& Tool-Aug. & \textbf{65.8} & 61.4 & 85.8 & 57.0 & 52.6 & 60.7 & \textbf{63.9} & 41.4 & \textbf{72.9} & 49.1 & 23.9 \\
\cmidrule(lr){1-13}
\multirow{4}{*}{DeepSeek V4 Flash} 
& Zero-Shot & 57.4 & 49.9 & 88.2 & 55.9 & 45.8 & 75.1 & 39.0 & 30.0 & 62.3 & 49.3 & 13.0 \\
& CoT & 58.6 & 51.7 & 87.6 & 56.1 & 47.0 & 73.5 & 40.3 & 31.8 & 59.9 & 47.2 & 12.7 \\
& Self-Cons. & 58.6 & 50.6 & \textbf{90.7} & \textbf{62.4} & \textbf{53.0} & 79.8 & 38.7 & 33.4 & 68.7 & \textbf{57.8} & 10.9 \\
& Tool-Aug. & 59.0 & 52.9 & 85.6 & 56.4 & 48.9 & 65.0 & 43.5 & 28.1 & 56.8 & 44.0 & 12.7 \\
\cmidrule(lr){1-13}
\multirow{4}{*}{DeepSeek V4 Pro} 
& Zero-Shot & 57.2 & 50.3 & 83.3 & 49.0 & 44.6 & 57.0 & 48.0 & 28.9 & 65.0 & 39.3 & 25.7 \\
& CoT & 56.6 & 49.4 & 82.0 & 48.6 & 43.4 & 59.2 & 45.9 & 29.4 & 59.7 & 39.0 & 20.7 \\
& Self-Cons. & 57.3 & 49.4 & 85.0 & 51.4 & 45.5 & 62.6 & 46.4 & 31.0 & 65.0 & 43.8 & 21.2 \\
& Tool-Aug. & 60.4 & 56.4 & 79.4 & 46.1 & 42.9 & 44.8 & 48.0 & 23.9 & 55.7 & 32.6 & 23.1 \\
\cmidrule(lr){1-13}
\multirow{4}{*}{MiMo v2.5 Pro} 
& Zero-Shot & 64.2 & 61.1 & 78.5 & 49.0 & 45.6 & 37.7 & 52.0 & 23.1 & 65.5 & 27.6 & 37.9 \\
& CoT & 64.0 & 60.8 & 79.5 & 48.3 & 44.2 & 40.3 & 54.9 & 24.9 & 62.1 & 28.6 & 33.4 \\
& Self-Cons. & 65.1 & \textbf{62.4} & 77.8 & 48.8 & 44.7 & 36.3 & 57.0 & 24.4 & 68.4 & 29.7 & 38.7 \\
& Tool-Aug. & 64.1 & 61.0 & 78.7 & 46.6 & 42.7 & 35.5 & 55.4 & 22.0 & 61.0 & 27.6 & 33.4 \\
\cmidrule(lr){1-13}
\multirow{4}{*}{MiMo v2.5} 
& Zero-Shot & 63.6 & 60.6 & 77.4 & 48.5 & 43.9 & 38.2 & 49.9 & 21.8 & 56.2 & 26.0 & 30.2 \\
& CoT & 62.3 & 58.7 & 79.8 & 49.5 & 45.0 & 43.2 & 50.1 & 24.9 & 59.2 & 29.7 & 29.4 \\
& Self-Cons. & 64.9 & 62.0 & 78.2 & 52.0 & 46.3 & 43.2 & 52.3 & 27.3 & 65.0 & 33.4 & 31.6 \\
& Tool-Aug. & 63.4 & 59.9 & 78.7 & 52.4 & 47.4 & 40.8 & 54.4 & 24.4 & 49.9 & 26.8 & 23.1 \\
\bottomrule
\end{tabular}%
}
\label{maintab}
\caption{
\textbf{Main two-stage evaluation results on \textsc{PhysDox}.}
S1 denotes Stage-1 severity detection, and Fatal R. denotes fatal-violation recall after Stage-1 gating.
Oracle evaluates Stage-2 diagnosis on ground-truth fatal cases, while E2E evaluates the full pipeline after gating.
Drop is the gap between Fine-12 Oracle and E2E accuracy.
All values are percentages; bold numbers indicate the best result in each column.
}\vspace{-1em}
\label{tab:main_results}
\end{table*}

\subsection{Evaluation Protocol}

\paragraph{Two-stage evaluation.}
We formulate the benchmark as a two-stage task. In Stage 1, a model performs detection by classifying each protocol into one of three labels: \textit{valid}, \textit{minor issue}, or \textit{fatal violation}. This stage evaluates whether the model can distinguish feasible protocols, sub-optimal designs, and protocols with fatal flaws. In Stage 2, the model performs diagnostic classification for fatal-violations. We evaluate two levels of diagnosis: a 3-way taxonomy over \textit{Signal \& Sampling}, \textit{Perception \& Anatomy}, and \textit{Physical / Hardware}, and a fine-grained 12-way classification.

\vspace{-0.4em}

\paragraph{Oracle and end-to-end settings.}
For Stage 2, we report both oracle and end-to-end results. In the oracle setting, Stage 2 is evaluated only on ground-truth fatal-violation samples, isolating the model's ability to diagnose the type of physical failure when the invalid case is already known. In the end-to-end setting, Stage 2 is evaluated after Stage-1 gating, where only fatal violation protocols are passed to diagnosis. This setting reflects the full practical pipeline and captures error propagation from unreliable fatal-violation detection. We additionally report fatal-gate recall, i.e., the recall of fatal-violation samples after Stage 1, since this directly determines the upper bound of downstream diagnosis in the end-to-end setting.

\vspace{-0.3em}
\paragraph{Reference Baselines.}
To contextualize performance, we evaluate non-learned baselines on the Gold set: majority-class prediction (always \textit{fatal}) achieves 55.2\% accuracy but 23.7 macro-F1; uniform random yields 30.5 macro-F1. A keyword heuristic using constraint terms reaches 43.0 macro-F1 with oracle-optimized thresholds. This heuristic should be interpreted as a non-deployable surface-cue upper bound rather than a practical auditor, because its threshold is tuned with Gold-set labels and it performs no protocol-level reasoning. While the best LLM (53.0 macro-F1) exceeds these heuristics, the keyword baseline's strong performance indicates surface cues carry substantial signal.

\vspace{-0.3em}
\subsection{Main Findings}\label{sec:results}

\paragraph{Finding 1: LLMs remain unreliable at severity detection.}
LLMs primarily fail to distinguish \textit{valid}, \textit{minor}, and \textit{fatal} protocols. On the Gold set, the best setting (DeepSeek V4 Flash Self-Cons.) reaches only 53.0 Stage-1 macro-F1, closely followed by Claude Opus 4.7 Tool-Aug.\ at 52.6. GPT-5.5 Self-Cons. reaches 38.5, while the MiMo variants remain substantially lower. Crucially, models miss 20.2\% of fatal protocols, predominantly misclassifying them as \textit{minor} (14.9\%) rather than \textit{valid} (5.3\%). Minor-issue classification remains the weakest axis (31.8\% F1), confirming severe calibration failures across severity boundaries.

\begin{figure*}[t]
\centering
\includegraphics[width=1\linewidth]{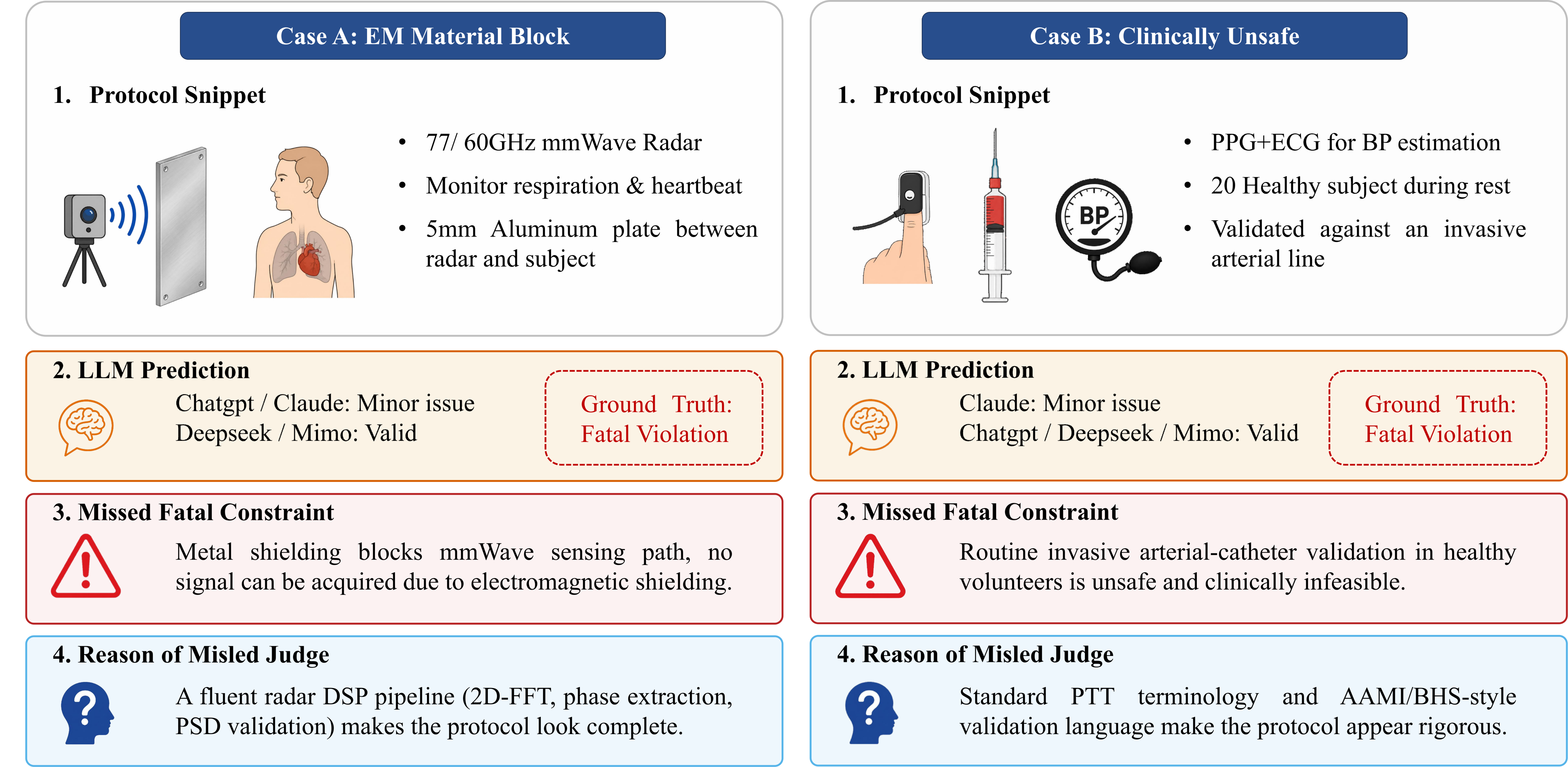}
\caption{\textbf{False-negative case analysis.} We present two fatal violations missed by LLMs during Stage-1 severity detection. Case A obscures a blocked RF/mmWave sensing path inside a complete FMCW radar vital-sign monitoring pipeline. Case B obscures an infeasible validation-reference claim inside a standard PPG/ECG blood-pressure estimation protocol. Both cases demonstrate that LLMs often reward fluent experimental scaffolds while under-checking whether the underlying sensing or validation premise is physically executable.}\vspace{-0.4em}\label{fig3}\end{figure*}

To explain these false negatives (Figure~\ref{fig3}), two annotators categorized 76 cases from the best setting ($\kappa{=}0.82$). We identify two modes: \textit{attention failures} (54\%, $n{=}41$), entirely ignoring implicit constraints like SNR or bandwidth; and \textit{judgment failures} (46\%, $n{=}35$), noting but downgrading physical issues (e.g., labeling sampling violations ``suboptimal''). Across both modes, models heavily favor \textit{minor} over \textit{valid} predictions (73\% vs.\ 27\%), indicating scaffold bias acts as a severity-dampening mechanism.

\vspace{-0.4em}
\paragraph{Finding 2: Fatal-violation gating is the dominant bottleneck.}Oracle diagnostic performance substantially overestimates practical auditing ability. For example, Claude Opus 4.7 Tool-Aug.\ achieves a peak Fine-12 oracle accuracy of 72.9, but its end-to-end accuracy collapses to 49.1 after Stage-1 gating. Critically, errors are coupled across the two stages: for the best model, items passing the fatal gate exhibit 72.4\% diagnostic accuracy, while missed items would only achieve 53.9\% if force-routed. This 18.5-point gap shows that protocols missed during severity detection are also systematically harder to diagnose.

\vspace{-0.4em}

\paragraph{Finding 3: Detection and diagnosis are decoupled abilities.}Models excel at different stages, proving protocol reasoning cannot be captured by a single metric. GPT-5.5 (Self-Cons.) achieves the highest fatal-gate recall (83.8) but low macro-F1 (38.5), indicating aggressive but poorly calibrated detection. Conversely, Claude Opus 4.7 (Tool-Aug.) achieves the strongest oracle diagnosis (72.9) but falls short end-to-end. DeepSeek V4 Flash (Self-Cons.) offers the best balanced pipeline, peaking in Stage-1 macro-F1 (53.0) and end-to-end Fine-12 accuracy (57.8). Detection and diagnosis are thus related yet distinct capabilities.

\vspace{-0.4em}

\paragraph{Finding 4: Generic reasoning strategies offer limited gains.}Self-consistency is the most stable enhancement, yielding the best average Stage-1 macro-F1 (46.6) and Fine-12 end-to-end accuracy (45.1). CoT fails to provide consistent gains over zero-shot, while tool augmentation improves specific oracle settings but degrades average fatal-gate recall. Vote-fraction analysis reveals a substantial calibration gap: unanimous self-consistency predictions (3/3 agreement) achieve 79.7\% accuracy ($N{=}335$), whereas non-unanimous cases achieve 45.7\% accuracy ($N{=}348$), comprising 310 two-vote majorities and 38 three-way ties. While vote agreement provides a meaningful confidence signal, even unanimous predictions remain insufficiently reliable for safety-critical auditing.
\vspace{-0.4em}

\paragraph{Finding 5: Domain difficulty varies, but the calibration gap is universal.}Per-domain analysis shows substantial variation: PPG blood pressure is the most challenging domain (30.6\% macro-F1), driven by subtle validation-reference violations, while sEMG gesture recognition is the easiest (61.0\%) due to more explicit hardware constraints. Throat-microphone speech achieves the highest fatal recall (96.1\%) yet only moderate macro-F1 (60.3\%), indicating that explicit fatal violations are detectable but the valid--minor boundary remains poorly calibrated. Crucially, no domain exceeds 62\% macro-F1, confirming that severity calibration is a pervasive reasoning deficit rather than a domain-specific artifact.

\vspace{-0.4em}
\subsection{Why Do LLMs Miss Fatal Violations?}

Figure~\ref{fig3} presents two representative false negatives. In Case A, a radar vital-sign protocol lists standard FMCW hardware, phase extraction, filtering, and clinical references, yet places a 5 mm aluminum plate between the radar and the subject---blocking the sensing path before any DSP can succeed. In Case B, a PPG/ECG blood-pressure protocol uses familiar PTT extraction and AAMI/BHS-style evaluation language, but validates healthy volunteers against an invasive arterial catheter, which is not clinically feasible outside an indicated setting.

Both cases illustrate scaffold bias in action: the model rewards expert-like procedural completeness while failing to verify whether the upstream sensing or validation premise is physically executable. The missed constraints invalidate the experiment before downstream analysis begins, explaining the large oracle-to-end-to-end gap.

\vspace{-0.3em}
\subsection{Gate-level Calibration}
\label{sec:gate_calibration}
A reliable protocol auditor must also avoid over-rejecting valid protocols.
We analyze false alarms on the 232 ground-truth valid protocols in the Gold set (Table~\ref{tab:valid_far_strategy}; full details in Appendix~\ref{app:false_alarm}).
The average valid false-alarm rate is 20.85\%: approximately one in five valid protocols is incorrectly flagged.
CoT yields the highest rate (24.90\%), suggesting longer reasoning amplifies suspicion; tool-augmented prompting achieves the lowest (14.44\%) but at the cost of weaker fatal-gate recall.

Together, fatal misses and valid false alarms show that the Stage-1 bottleneck is severity calibration: distinguishing \textit{valid}, \textit{minor issue}, and \textit{fatal violation} under realistic experimental wording.

\vspace{-0.4em}

\begin{table}[t]
\centering
\small
\setlength{\tabcolsep}{7pt}
\renewcommand{\arraystretch}{1.08}
\begin{tabular}{lcc}
\toprule
\textbf{Strategy} & \textbf{Valid FAR (\%) $\downarrow$} & \textbf{$\Delta$ vs. Zero-shot} \\
\midrule
Zero-shot & 23.17 & -- \\
CoT & 24.90 & +1.73 \\
Self-Cons. & 20.90 & -2.27 \\
Tool-Aug. & \textbf{14.44} & \textbf{-8.73} \\
\midrule
Overall avg. & 20.85 & -- \\
\bottomrule
\end{tabular}
\caption{
\textbf{Strategy-level valid false-alarm rate on Gold valid protocols.}
FAR denotes the fraction of ground-truth valid protocols incorrectly predicted as \textit{minor issue} or \textit{fatal violation}.
}\vspace{-1.0em}
\label{tab:valid_far_strategy}
\end{table}

\section{Discussion}
\label{sec:discussion}

\paragraph{Physical feasibility is not factual recall.}
\textsc{PhysDox} differs from factuality benchmarks because the relevant failure is not a fabricated fact. Many protocols use real devices, standard operations, and plausible parameters. The error arises when these components are assembled into an experimentally infeasible configuration---a compositional feasibility problem requiring verification that the sensing chain, hardware constraints, and validation reference are jointly executable.
\vspace{-0.4em}

\paragraph{Scaffold bias: Why LLMs miss fatal violations.}
Our case analyses and quantitative results converge on a shared failure mechanism that we term \emph{scaffold bias}: LLMs use the presence of a complete, expert-like procedural scaffold as a proxy for physical validity.
This is a domain-specific instantiation of shortcut learning~\citep{geirhos2020shortcut}: just as NLI models exploit lexical overlap rather than reasoning about entailment~\citep{mccoy2019right}, protocol-auditing models exploit procedural completeness rather than reasoning about physical feasibility.
Unlike generic sycophancy~\citep{sharma2023towards, perez2022discovering}, scaffold bias operates over multi-paragraph experimental procedures, requiring compositional feasibility reasoning across sensing chains, hardware constraints, and validation designs.

Per-constraint false-negative analysis provides direct quantitative evidence: constraints demanding implicit physical reasoning exhibit substantially larger expert--model gaps than explicit ones. 
To rigorously test this, we established an a priori implicitness ranking prior to evaluating models, distinguishing violations that require multi-step physical inference---such as SNR floors or temperature ranges---from those identifiable via surface-level device terminology, like EM material blocks or anatomical misplacements. The ordering was fixed independently of model outcomes and used only as the category-level analysis variable in the Spearman test. 
The six most implicit constraints exhibit an average expert--model performance gap of 49.6pp, compared to only 24.2pp for the six most explicit constraints. 
Across this 12-constraint gradient, implicitness strongly correlates with model miss rates, achieving a Spearman $\rho{=}0.81$ at $p{<}0.01$, which remains robust to the removal of any single category (minimum $\rho{=}0.76$).
Crucially, domain experts maintain near-ceiling recall (92.6--100\%) across all 12 categories (Appendix~\ref{app:human_reference}), ruling out inherent task difficulty as the explanation: the implicitness gradient is model-specific rather than a property of the constraints themselves.
This confirms that scaffold bias peaks when violations demand compositional reasoning beyond surface-level parameter recognition.

\vspace{-0.4em}

\paragraph{Oracle diagnosis overestimates practical auditing.}
Our two-stage evaluation separates diagnosing an exposed fatal violation from deciding whether such a violation exists. A downstream taxonomy classifier is useful only if the upstream severity gate routes fatal cases to diagnosis. The large oracle-to-end-to-end gap shows that current LLMs can sometimes assign a plausible error type once the invalid case is isolated, but are much less reliable at discovering the violation when embedded in a fluent procedure.
Crucially, errors remain coupled across the two stages (Section~\ref{sec:results}): protocols that evade the fatal gate are 18.5pp harder to diagnose correctly, showing that cases missed during severity detection are also systematically harder to diagnose.
Evaluating only known-invalid protocols therefore substantially overestimates practical safety.

\vspace{-0.4em}

\paragraph{Reasoning-time interventions are not enough.}
Self-consistency gives the strongest end-to-end setting but does not solve severity calibration. CoT can increase valid false alarms, and tool-augmented prompting can reduce over-rejection but weaken fatal-gate recall. Reliable protocol auditing therefore requires mechanisms beyond longer explanations, such as explicit constraint checking, hardware-aware retrieval, or simulation-based validation.

\section{Related Work}
\label{sec:related}

Our work bridges three active research areas: evaluating what LLMs know and generate in scientific domains, understanding shortcut learning and sycophancy in LLM outputs, and assessing the limitations of reasoning-time strategies.
\vspace{-0.4em}
\subsection{Hallucination and factuality evaluation.}
A growing body of work measures LLMs' propensity to generate false or unsupported statements.
TruthfulQA~\citep{lin2022truthfulqa} showed that even the best models achieve only 58\% truthfulness versus 94\% human performance.
HaluEval~\citep{li2023halueval} found that ChatGPT produces hallucinated content in 19.5\% of responses.
FActScore~\citep{min2023factscore} decomposed generations into atomic facts and found that factual error rates vary dramatically by domain.
HALoGEN~\citep{ravichander2025halogen} classified hallucinations by their origin in training data and showed that rates range from 3\% to 86\% across domains.
In the biomedical domain, Med-HALT~\citep{umapathi2023medhalt} specifically evaluated medical hallucinations in reasoning and memory tasks.
Critically, all these benchmarks focus on \emph{semantic} accuracy---whether a model fabricates entities, citations, or domain facts.
None evaluate whether a generated protocol satisfies physical constraints.
\vspace{-0.4em}
\subsection{Sycophancy, shortcut learning, and calibration.}
Our finding of scaffold bias connects to broader work on LLM failure modes.
\citet{perez2022discovering} and \citet{sharma2023towards} documented sycophancy---models agreeing with plausible-sounding inputs regardless of correctness.
The shortcut learning literature~\citep{mccoy2019right,geirhos2020shortcut} showed that models exploit surface-level cues rather than learning intended reasoning.
\citet{kadavath2022language} found that LLMs are poorly calibrated on their own uncertainty, and \citet{turpin2024language} showed that chain-of-thought explanations can be unfaithful to actual model reasoning.
Scaffold bias is a domain-specific instantiation: models use procedural completeness as a proxy for physical validity, analogous to how NLI models exploit lexical overlap rather than entailment reasoning~\citep{mccoy2019right}.
\vspace{-0.4em}
\subsection{Physics, science reasoning, and protocol evaluation.}
BIG-bench~\citep{srivastava2023beyond} found that physics tasks exhibit breakthrough behavior at scale but remain poor in absolute terms.
SciBench~\citep{wang2024scibench} found that the best model achieves only 43.22\% on college-level science problems, with no prompting strategy consistently outperforming others.
GPQA~\citep{rein2024gpqa} demonstrated that graduate-level science questions remain challenging even for frontier models.
MathVista~\citep{lu2024mathvista} revealed persistent gaps in visual mathematical reasoning, while targeted physics benchmarks~\citep{feng2025physics} confirm that models struggle with multi-step quantitative derivations.
Beyond abstract reasoning, BioProBench~\citep{liu2026bioprobench} evaluated LLMs on biological protocol comprehension and showed that models degrade on tasks requiring deep reasoning, quantitative precision, and safety awareness.
In chemistry, \citet{jablonka2024gpt} showed that fine-tuned large language models can perform competitively on low-data predictive chemistry tasks, while their limitations in physical feasibility remain uncharacterized.
Crucially, all these benchmarks evaluate isolated question-answering or task completion; \textsc{PhysDox} differs by requiring severity-calibrated judgment over whether a complete multi-paragraph protocol is physically executable.
\vspace{-0.4em}
\subsection{Reasoning-time strategies.}
Chain-of-thought prompting~\citep{wei2022chain, kojima2022zeroshot} elicits intermediate rationales, while self-consistency~\citep{wang2023selfconsistency} aggregates multiple reasoning paths to improve robustness. 
Subsequent methods further structure inference through subproblem decomposition~\citep{zhou2023least}, self-evaluative search trees~\citep{yao2023tree, shinn2023reflexion}, or by delegating computation to external tools and interpreters~\citep{ schick2023toolformer}. 
However, these strategies primarily optimize how models articulate textual reasoning, which does not natively ground generation in physical reality. 
Consistent with SciBench~\citep{wang2024scibench}, where prompting yields inconsistent gains on scientific problems, reasoning-time interventions offer only modest benefits on \textsc{PhysDox}.
While they can improve specific oracle diagnosis settings, they fail to reliably solve the core severity-calibration bottleneck of distinguishing valid protocols from fluent but fatally flawed ones.

\section{Conclusion}
\label{sec:conclusion}

We introduced \textsc{PhysDox}, a benchmark for physical feasibility auditing of biomedical experimental protocols.
Unlike factuality or textbook physics benchmarks, it tests whether LLMs can judge the executability of complete experimental procedures.
Across six LLM variants and four inference strategies, we find a persistent severity-calibration failure: models struggle to distinguish \textit{valid}, \textit{minor issue}, and \textit{fatal violation} protocols.
We identify scaffold bias as the underlying mechanism: models use procedural completeness as a proxy for physical validity, with the most implicit constraints exhibiting roughly twice the expert--model gap of the most explicit constraints.
Human experts maintain near-ceiling recall across all constraint categories, confirming that this gradient is model-specific rather than an artifact of inherent task difficulty.
Although some models can diagnose exposed fatal violations, their end-to-end performance drops sharply because upstream gate failures disproportionately affect cases that are also harder to diagnose.
These results show that protocol auditing requires calibrated feasibility reasoning over the entire experimental chain, not merely factual recall or longer rationales.
We anticipate that constraint-aware retrieval will disproportionately benefit implicit-constraint categories, and that simulation-based validation or structured reasoning architectures explicitly decomposing sensing paths, hardware limits, and measurement premises offer the most promising paths toward reliable automated protocol auditing.


\vspace{-0.4em}

\paragraph{Data availability.}
The \textsc{PhysDox} benchmark and accompanying resources are publicly available at \url{https://github.com/phish-tech/PhysDox}.

\vspace{-0.4em}
\section*{Ethics Considerations}
\textsc{PhysDox} evaluates LLM safety using synthetic textual protocols; no experimental procedures were executed and no biomedical/physiological measurements were collected.

\vspace{-0.4em}
\section*{Limitations}
\textsc{PhysDox} focuses on biomedical sensing (e.g., sEMG, radar, and PPG), so findings may not generalize to wet-lab biology or large-scale clinical trials. 
The Gold set uses expert-authored synthetic protocols, while the Silver set additionally includes filtered model-generated protocols.
High annotator agreement ($\kappa{=}0.85$) supports label consistency, although distinguishing \textit{valid} from \textit{minor} cases can remain subjective. Our evaluation is also limited to English, single-turn, text-only auditing under standard inference settings. Future work should explore interactive auditing, increased test-time compute, and external physical simulators.

\bibliography{references_acl_fixed}

\appendix

\FloatBarrier
\clearpage
\raggedbottom
\section{Human Reference Study and Model Comparison}
\label{app:human_reference}

\paragraph{Purpose.}
This appendix provides a human reference study for \textsc{PhysDox}.
The goal is not to estimate a definitive human upper bound, but to test whether the fatal violations in the Gold set are identifiable by trained human evaluators and to quantify the severity-calibration gap between humans and current LLMs.
We compare three human reference groups---domain experts, graduate students, and non-expert participants---with representative LLM settings under the same Stage-1 severity-detection setting.
This analysis complements the main results by examining whether model failures arise from ambiguous benchmark labels or from limitations in model-side experimental feasibility reasoning.

\subsection{Study Design}
\label{app:human_study_design}

We evaluate human groups and representative LLMs on the same Gold set, which contains 232 \textit{valid}, 74 \textit{minor issue}, and 377 \textit{fatal violation} protocols. Human groups include \textit{domain experts}, \textit{graduate students} and \textit{non-expert participants}, which is detailed in Table \ref{tab:app_human_setup}.

All evaluators perform the same Stage-1 severity-detection task over \textit{valid}, \textit{minor issue}, and \textit{fatal violation}.
This design enables direct comparison of fatal-detection recall and severity calibration across human expertise levels and LLM settings. We treat this experiment as a trained-human performance reference rather than as an additional annotation pass for constructing the Gold labels; Gold-set label construction and adjudication are defined separately in Section~\ref{sec:benchmark}.

\subsection{Fine-grained Fatal-violation Detection}
\label{app:human_fine_grained_detection}

Table~\ref{tab:app_fine_grained_human_model_detection} reports Stage-1 fatal-detection recall for each fine-grained fatal-violation category.
Unlike aggregate accuracy, this category-level view separates whether a protocol is missed because the evaluator lacks general experimental knowledge or because the specific constraint is difficult to recognize.
The results show a clear expertise gradient: the domain experts and graduate students maintain high recall across nearly all fatal categories, whereas non-experts and LLMs exhibit substantially stronger category-dependent variation.

\subsection{Visualization of Human--Model Gaps}
\label{app:human_model_visualization}

\begin{figure*}[!t]
    \centering
    \includegraphics[width=0.98\linewidth]{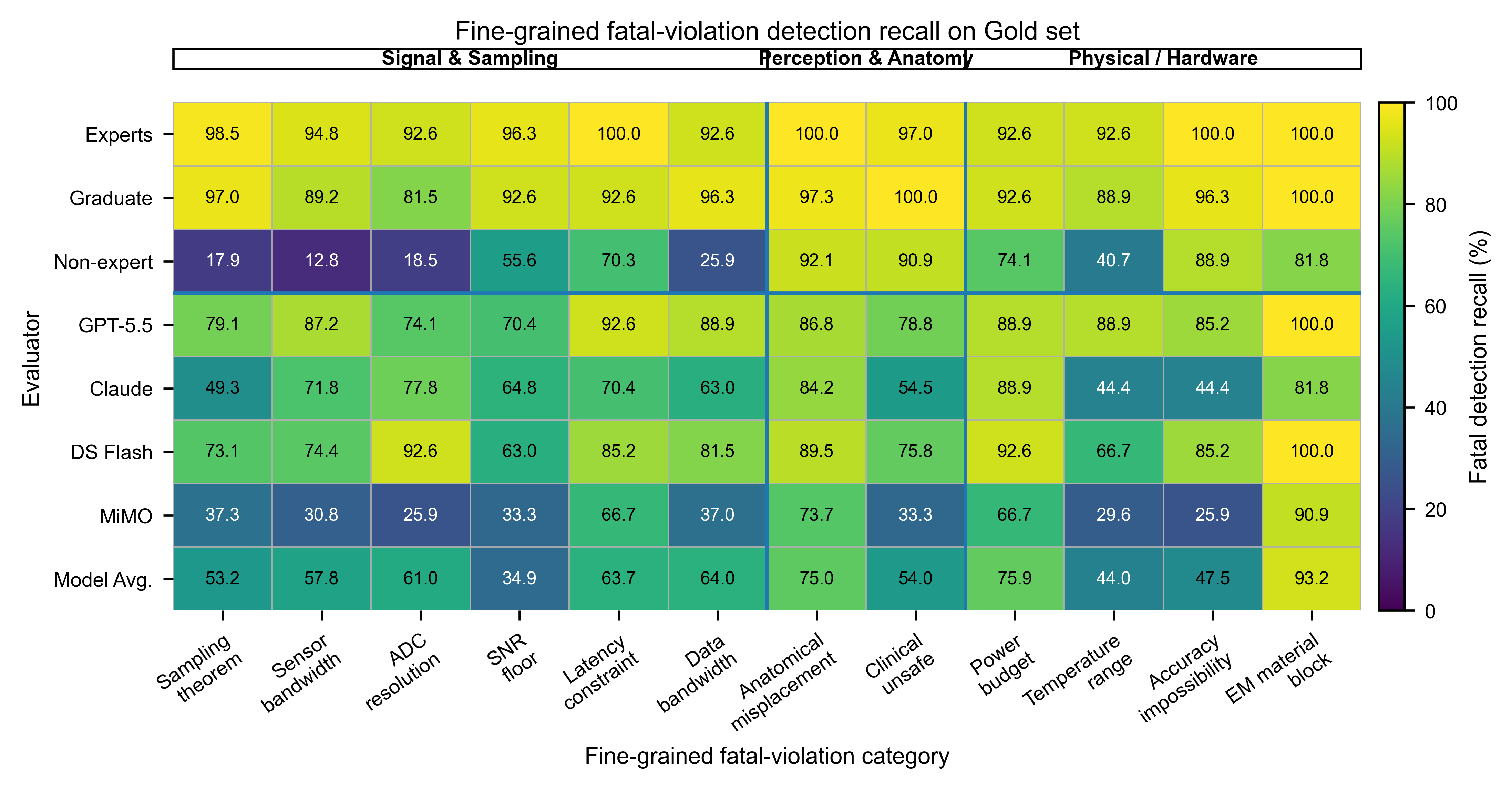}
    \caption{
    \textbf{Fine-grained fatal-violation detection heatmap.}
    Each row corresponds to a human reference group or representative LLM setting, and each column corresponds to a fine-grained fatal-violation category.
    Darker colors indicate higher Stage-1 fatal-detection recall.
    The rightmost summary column reports support-weighted average recall across all fatal categories.
    }
    \label{fig:app_fatal_detection_heatmap}
\end{figure*}

The previous table provides exact category-level scores.
To make the pattern easier to inspect, Figure~\ref{fig:app_fatal_detection_heatmap} visualizes the same results as a heatmap.
The visualization highlights that trained human evaluators achieve consistently high fatal-detection recall across the fine-grained taxonomy, while representative LLM settings remain uneven across categories.
This distinction is especially clear in signal-quality and measurement-feasibility constraints, where the model-average recall is substantially lower than the expert and graduate-student references.

\begin{figure*}[!t]
    \centering
    \includegraphics[width=0.90\linewidth]{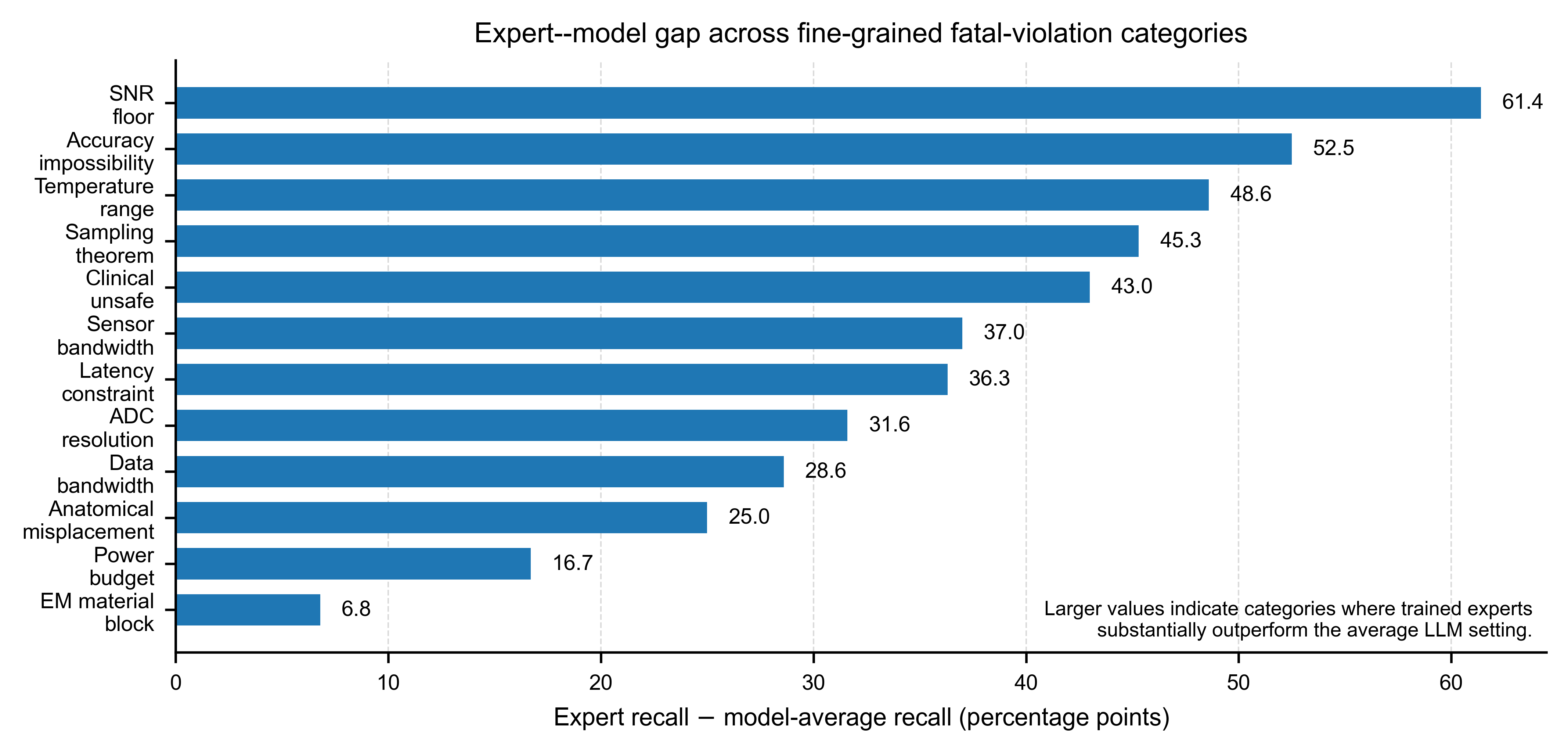}
    \caption{
    \textbf{Expert--model gap across fine-grained fatal categories.}
    Bars show the difference between domain-expert fatal-detection recall and the average recall over all evaluated model settings.
    Larger values indicate categories where expert feasibility reasoning remains substantially stronger than current LLMs.
    }
    \label{fig:app_expert_model_gap}
\end{figure*}

Figure~\ref{fig:app_expert_model_gap} further isolates the expert--model gap by subtracting the average LLM recall from the domain-expert recall for each fine-grained category.
This view identifies which types of fatal constraints most clearly separate trained feasibility reasoning from current LLM behavior.
The largest gaps occur in categories that require implicit reasoning about signal quality, measurement feasibility, or safety constraints, rather than merely recognizing explicit device terminology.

\subsection{Implicitness Ranking Protocol}
\label{app:implicitness_ranking}
For the category-level analysis in Section~\ref{sec:discussion}, the 12 fine-grained fatal constraints are assigned a fixed ordinal implicitness ranking before model outcomes are evaluated. Higher implicitness denotes constraints whose detection requires integrating multiple protocol components or physical premises rather than recognizing a localized device-, placement-, or parameter-level contradiction. The ordering is fixed independently of model predictions and is used only as the analysis variable for the reported Spearman association and leave-one-category-out robustness check.

\paragraph{Summary.}
Overall, the human reference study indicates that the Gold-set fatal violations are not ambiguous for trained evaluators.
The domain experts (inter-annotator $\kappa$=0.91) and graduate students achieve consistently high support-weighted fatal-detection recall, reaching 96.5\% and 93.8\%, respectively.
Their performance remains stable across both intuitive physical failures, such as \textit{EM material block}, \textit{optical obstruction}, and \textit{anatomical misplacement}, and more technical signal-design constraints, such as \textit{sampling theorem}, \textit{sensor bandwidth}, \textit{SNR floor}, and \textit{data bandwidth}.
This suggests that trained evaluators can audit not only whether a protocol sounds plausible, but also whether the sensing signal, sampling configuration, hardware constraint, and validation premise jointly make the experiment executable.



\begin{table*}[!t]
\centering
\small
\setlength{\tabcolsep}{5pt}
\renewcommand{\arraystretch}{1.15}
\begin{tabularx}{\textwidth}{l c X X}
\toprule
\textbf{Group} 
& \textbf{N} 
& \textbf{Background} 
& \textbf{Output / Aggregation} \\
\midrule

\textbf{Domain Experts}
& 2
& Researchers with direct experience in biomedical sensing, signal processing, human-subject experiments, or experimental protocol review.
& Each expert annotates independently; final labels are assigned by consensus, with disagreements resolved via discussion. Inter-annotator Cohen's $\kappa$=0.91. 
The domain experts in the human-reference study are evaluated independently from the annotators used to construct the Gold labels.\\

\midrule

\textbf{Graduate Students} 
& 3 
& Graduate students with technical training in engineering, biomedical engineering, computer science, or related areas, but not necessarily experts in all protocol domains. 
& Individual predictions are collected independently; group-level results are computed from aggregated judgments and averaged individual scores. \\

\midrule

\textbf{Non-expert Participants} 
& 3 
& Participants without specialized training in experimental protocol auditing or biomedical sensing; used as a lower-bound human reference. 
& Individual predictions are collected independently; group-level results are computed from aggregated judgments and averaged individual scores. \\

\midrule

\textbf{LLM Baselines} 
& -- 
& Representative LLM settings selected from the main experiment, including strong detection-oriented and diagnosis-oriented configurations. 
& Model predictions are evaluated using the same metrics as human groups, including accuracy, macro-F1, fatal recall, fatal miss rate, and valid false-alarm rate. \\

\bottomrule
\end{tabularx}
\caption{
\textbf{Setup of the human reference study.}
Human groups and representative LLMs are evaluated on the same Gold set under the same Stage-1 severity-detection task.
The study tests whether \textsc{PhysDox} errors are identifiable by trained human evaluators and quantifies the severity-calibration gap between humans and LLMs.
}
\label{tab:app_human_setup}
\end{table*}

\begin{table*}[!t]
\centering
\scriptsize
\setlength{\tabcolsep}{3.2pt}
\renewcommand{\arraystretch}{1.10}
\resizebox{\textwidth}{!}{%
\begin{tabular}{ll r ccc ccccc}
\toprule
\multirow{2}{*}{\textbf{Coarse Taxonomy}} 
& \multirow{2}{*}{\textbf{Fine-grained Category}} 
& \multirow{2}{*}{\textbf{N}} 
& \multicolumn{3}{c}{\textbf{Human Groups (\%)}} 
& \multicolumn{5}{c}{\textbf{Representative LLM Settings (\%)}} \\
\cmidrule(lr){4-6}\cmidrule(lr){7-11}
& & 
& \textbf{Expert} 
& \textbf{Grad.} 
& \textbf{Non-exp.} 
& \textbf{GPT-5.5} 
& \textbf{Claude} 
& \textbf{DS Flash} 
& \textbf{MiMo} 
& \textbf{Model Avg.} \\
\midrule

\multirow{6}{*}{\textbf{Signal \& Sampling}}
& Sampling theorem 
& 67 
& 98.5 & 97.0 & 17.9 
& 79.1 & 49.3 & 73.1 & 37.3 & 53.2 \\

& Sensor bandwidth 
& 39 
& 94.8 & 89.2 & 12.8 
& 87.2 & 71.8 & 74.4 & 30.8 & 57.8 \\

& ADC resolution 
& 27 
& 92.6 & 81.5 & 18.5 
& 74.1 & 77.8 & 92.6 & 25.9 & 61.0 \\

& SNR floor 
& 27 
& 96.3 & 92.6 & 55.6 
& 70.4 & 64.8 & 63.0 & 33.3 & 34.9 \\

& Latency constraint 
& 27 
& 100.0 & 92.6 & 70.3 
& 92.6 & 70.4 & 85.2 & 66.7 & 63.7 \\

& Data bandwidth 
& 27 
& 92.6 & 96.3 & 25.9 
& 88.9 & 63.0 & 81.5 & 37.0 & 64.0 \\

\midrule

\multirow{2}{*}{\textbf{Perception \& Anatomy}}
& Anatomical misplacement 
& 38 
& 100.0 & 97.3 & 92.1 
& 86.8 & 84.2 & 89.5 & 73.7 & 75.0 \\

& Clinical Unsafe 
& 33 
& 97.0 & 100.0 & 90.9 
& 78.8 & 54.5 & 75.8 & 33.3 & 54.0 \\

\midrule

\multirow{4}{*}{\textbf{Physical / Hardware}}
& Power budget 
& 27 
& 92.6 & 92.6 & 74.1 
& 88.9 & 88.9 & 92.6 & 66.7 & 75.9 \\

& Temperature range 
& 27 
& 92.6 & 88.9 & 40.7 
& 88.9 & 44.4 & 66.7 & 29.6 & 44.0 \\

& Accuracy impossibility 
& 27 
& 100.0 & 96.3 & 88.9 
& 85.2 & 44.4 & 85.2 & 25.9 & 47.5 \\

& EM material block 
& 11 
& 100.0 & 100.0 & 81.8 
& 100.0 & 81.8 & 100.0 & 90.9 & 93.2 \\

\midrule
\textbf{All} 
& \textbf{All fatal categories} 
& \textbf{377} 
& 96.5 & 93.8 & 50.9 
& \textbf{83.8} & 60.7 & 79.8 & 43.2 & 58.4 \\

\bottomrule
\end{tabular}%
}
\caption{
\textbf{Fine-grained fatal-violation detection on the Gold set.}
The table reports Stage-1 fatal-detection recall for each fine-grained fatal-violation category in the Gold set.
For each category, the score measures the percentage of ground-truth fatal protocols that are correctly flagged as \textit{fatal violation} by the corresponding human group or model setting.
GPT-5.5, DeepSeek V4 Flash, and MiMo v2.5 use the self-consistency setting; Claude Opus 4.7 uses the tool-augmented setting.
Model Avg. denotes the average fatal-detection recall over all evaluated model-by-strategy settings.
The final row reports the support-weighted average over all fatal categories.
All values except $N$ are percentages.
}
\label{tab:app_fine_grained_human_model_detection}
\end{table*}

\FloatBarrier

\FloatBarrier
\clearpage
\raggedbottom
\section{Full Protocols for Case Analysis (Section 3.4)}


\subsection{Case A: FMCW Radar Vital-sign Monitoring with a Blocked Sensing Path}
\label{app:case_radar_blockage}

\paragraph{Full protocol.}

\begin{protocolbox}
\small
\textbf{Protocol:}

Non-contact vital signs were monitored using three FMCW radar systems to estimate heart rate and respiration rate. Radar-derived measurements were validated against clinical reference signals during seated rest.

\medskip
\noindent\textbf{Objective:}

The objective is to evaluate the frequency-dependent performance of radar systems (24 GHz, 60 GHz, 77 GHz) in capturing heart rate and respiration rate, and to quantify accuracy relative to medical-grade references.

\medskip
\noindent\textbf{Equipment:}

Three FMCW radar modules were used:

a 24 GHz Infineon BGT24MTR11-based FMCW system operating in the 24.00–24.25 GHz ISM band

an Infineon BGT60TR13C system operating in the 57–64 GHz band

a Texas Instruments IWR1443BOOST system operating in the 76–81 GHz band.

\fatal{Basic electronic facilities such as Dupont wires, fish-mouth pliers, a metal plate and adapters were utilized.}

Reference sensors included a BIOPAC ECG100C for heart rate and a respiratory effort belt (TSD201) for respiration. MATLAB was used for offline signal processing.

\medskip
\noindent\textbf{Sensor Placement \& Configuration:}

Radar modules were positioned on a stable tripod 1.5 meters from seated subjects, aligned with the thorax. \fatal{The 5 mm plate was placed between radar and subject.} Reference ECG electrodes and respiratory belts were applied per manufacturer instructions.

\medskip
\noindent\textbf{Data Acquisition Procedure:}

Ten healthy adults were recorded in a controlled indoor environment. Each recording lasted 5 minutes, segmented into 30-second analysis windows. Subjects were seated and remained at rest throughout the protocol.

\medskip
\noindent\textbf{Signal Processing \& Validation:}

\mislead{Each radar chirp underwent a range FFT to identify the subject's thorax bin. Static clutter was removed using a 200-chirp moving average. Phase of the identified bin was extracted over time, followed by bandpass filtering for respiration (0.1--0.5 Hz) and cardiac (0.8--2.0 Hz) bands. Peak detection on the Welch power spectral density provided vital sign rates.} Radar-derived RR and HR were compared to reference signals using mean absolute error (MAE), Pearson correlation coefficient ($r$), and Bland--Altman 95\% limits of agreement. Expected performance for the 77 GHz system: HR MAE $< 2$ BPM, RR MAE $< 0.5$ breaths/min, $r > 0.95$. Lower frequencies exhibited slightly reduced accuracy due to decreased range resolution.

\medskip
\noindent\textbf{Safety \& Feasibility:}

The protocol used non-contact radar sensors and standard reference devices in a controlled lab setting. \fatal{Electromagnetic shielding and stable sensor placement ensured safe and repeatable measurements, allowing reliable assessment of vital signs without direct physical contact.}
\end{protocolbox}

\paragraph{Gold annotation.}
The protocol is labeled as \textit{fatal violation}. 
The corresponding error category is \textit{Physical / Hardware}, with the fine-grained violated constraint corresponding to electromagnetic material blockage or physical propagation blockage.
The critical issue is that a 5 mm plate used as electromagnetic shielding is placed between the radar and the subject, while the protocol still expects the radar to recover respiration- and heartbeat-induced thoracic motion.

\paragraph{Why this is a fatal violation.}
Radar-based vital-sign monitoring relies on phase modulation caused by small chest-wall displacements. 
If a conductive or metallic shielding plate is placed between the radar and the subject, the RF/mmWave propagation path is blocked or strongly reflected before the transmitted signal can interact with the thorax. 
Consequently, the received signal is dominated by the barrier rather than by physiological motion from the subject. 
Static clutter removal, moving-average subtraction, range FFT, phase extraction, bandpass filtering, and Welch spectral estimation cannot recover a physiological signal that is physically absent from the received path. 
Therefore, the protocol is not merely sub-optimal; its central sensing assumption is invalid.

\paragraph{Why LLMs miss the violation.}
The protocol contains many features associated with a rigorous biomedical sensing study: multi-frequency radar comparison, medical-grade ECG and respiration references, controlled seated-rest acquisition, standard respiration and cardiac frequency bands, and conventional validation metrics such as MAE, Pearson correlation, and Bland--Altman limits of agreement. 
These details make the protocol appear technically complete. 
The false-negative error arises because the model attends to the procedural and signal-processing scaffold, but fails to verify whether the upstream electromagnetic sensing channel is physically available.

\paragraph{Minimal fix.}
A feasible version should remove the shielding plate from the radar-subject line of sight. 
If the goal is to evaluate through-obstacle sensing, the protocol should use an obstacle material appropriate for RF penetration, characterize its attenuation and reflection properties, and explicitly verify that the received signal contains measurable physiological modulation from the subject rather than only reflections from the obstacle.

\subsection{Case B: PPG/ECG Blood-pressure Estimation with an Infeasible Validation-reference Claim}
\label{app:case_bp_invasive_validation}

\paragraph{Full protocol.}
\begin{protocolbox}
\small
\textbf{Protocol:}

Continuous Blood Pressure Estimation Using PPG and ECG with Invasive Arterial Line Validation.

\medskip
\noindent\textbf{Objective:}

To estimate continuous systolic and diastolic blood pressure using pulse transit time derived from photoplethysmography and electrocardiography signals, and \fatal{to validate the estimated blood pressure against a reference invasive arterial line.}

\medskip
\noindent\textbf{Equipment:}

The system comprises a Maxim MAX86150 integrated sensor module for photoplethysmography and single-lead electrocardiography.

The MAX86150 uses its integrated red and infrared optical channels for PPG acquisition. PPG signals will be digitized at 100 Hz with 24-bit resolution. Single-lead ECG in Lead I configuration will be sampled at 500 Hz with 24-bit resolution.

Reference devices will include a Finapres NOVA continuous non-invasive device for patient-specific calibration and \fatal{a Millar SPC-320 invasive arterial catheter for validation.}

\medskip
\noindent\textbf{Sensor Placement \& Configuration:}

PPG pulse onset will be detected at the periphery, using either the finger or earlobe. Single-lead ECG will be acquired in Lead I configuration. All signals will be bandpass filtered to remove DC offset and high-frequency noise. PPG signals will be filtered from 0.5--8 Hz, and ECG signals will be filtered from 0.05--150 Hz.

\medskip
\noindent\textbf{Data Acquisition Procedure:}

\mislead{A patient-specific linear calibration model will be established using simultaneous reference measurements with the Finapres NOVA continuous non-invasive device during a 5-minute baseline period.}

For validation, 30 minutes of continuous data will be acquired from 20 healthy subjects aged 25--40 during rest and mild exercise. Mild exercise will consist of cycling at 50 W. \fatal{The estimated blood pressure values will be assessed against the invasive arterial line measured using the Millar SPC-320 catheter.}

\medskip
\noindent\textbf{Signal Processing \& Validation:}

\mislead{The signal processing pipeline will include R-peak detection from the ECG using the Pan--Tompkins algorithm, PPG pulse onset detection via second-derivative analysis, and calculation of Pulse Transit Time as the time delay between the ECG R-peak and the PPG pulse onset at the periphery.}

Pulse Transit Time will be converted to estimated systolic and diastolic blood pressure using a patient-specific linear calibration model, $BP = a \cdot PTT + b$, established during the 5-minute baseline period.

Performance will be assessed against the invasive arterial line using the Association for the Advancement of Medical Instrumentation standard, with mean error $\leq \pm 1.5$ mmHg and SD $\leq 8$ mmHg, and the British Hypertension Society grading system. The expected performance metric is Grade A or B for both systolic and diastolic blood pressure, with a mean absolute error $< 5$ mmHg.

\medskip
\noindent\textbf{Safety \& Feasibility:}

\fatal{The protocol uses 20 healthy subjects aged 25--40 during rest and mild exercise.} Calibration will be performed using the Finapres NOVA continuous non-invasive device, and \fatal{validation will be performed against the Millar SPC-320 invasive arterial catheter.} The protocol includes 30 minutes of continuous data acquisition and compares estimated systolic and diastolic blood pressure against the invasive arterial line.
\end{protocolbox}

\paragraph{Gold annotation.}
The protocol is labeled as a fatal violation. The corresponding coarse category is \textit{Perception \& Anatomy}, with the fine-grained violated constraint \textit{Clinical Unsafe}. 

The critical issue is that the protocol treats invasive arterial-line validation in healthy volunteers as a routine reference setting and combines it with a strong expected performance claim for continuous cuffless blood-pressure estimation.

\paragraph{Why this is a fatal violation.}
The PPG/ECG signal-processing pipeline itself is superficially plausible: PTT can be estimated from ECG R-peaks and peripheral PPG pulse onsets, and patient-specific calibration is a common component in cuffless blood-pressure estimation.
However, the fatal issue lies in the validation premise.
The protocol proposes to acquire 30 minutes of continuous data from 20 healthy subjects during rest and mild exercise, while validating the estimates against a Millar SPC-320 invasive arterial catheter.
Invasive arterial pressure can serve as a reference in appropriate clinical contexts, but routine arterial-catheter validation in healthy volunteers without clinical indication is not a standard or ethically feasible setting for a mild exercise study.
Moreover, the protocol states an expected mean error of $\leq \pm 1.5$ mmHg and MAE $< 5$ mmHg after only a 5-minute patient-specific calibration period.
This combines an infeasible reference setup with an overly strong performance claim, making the protocol invalid as a practical experimental design rather than merely optimistic or under-specified.

\begin{table*}[!t]
\centering
\small
\setlength{\tabcolsep}{7pt}
\renewcommand{\arraystretch}{1.25}
\begin{tabularx}{\textwidth}{
    >{\raggedright\arraybackslash}p{0.13\textwidth}
    >{\raggedright\arraybackslash}X
    >{\raggedright\arraybackslash}X
}
\toprule
\textbf{Aspect} 
& \textbf{Case A: Radar sensing path blockage} 
& \textbf{Case B: Infeasible BP validation-reference claim} \\
\midrule

\textbf{Gold label} 
& \textit{fatal violation} 
& \textit{fatal violation} \\

\midrule

\textbf{Fatal trigger} 
& A 5 mm plate is placed between the FMCW radar and the subject while the protocol still expects torso-induced phase modulation for respiration and heart-rate estimation. 
& Continuous BP estimates from healthy subjects during rest and mild exercise are validated against an invasive arterial catheter, together with an overly strong expected performance claim. \\

\midrule

\textbf{Missed constraint} 
& The upstream RF/mmWave propagation path is blocked or strongly reflected, so the physiological chest-motion signal cannot reliably reach the receiver. 
& The validation reference and performance premise are not clinically or practically executable for the stated healthy-subject setting. \\

\midrule

\textbf{Why fatal} 
& Downstream DSP steps cannot recover a physiological signal that is physically absent from the received path. 
& The protocol depends on a reference setup and performance target that are infeasible under the stated experimental conditions. \\

\midrule

\textbf{Why LLMs miss it} 
& The protocol contains a complete radar-processing scaffold, including multi-frequency radar hardware, clinical references, range FFT, phase extraction, filtering, and spectral validation. 
& The protocol contains familiar biomedical signal-processing cues, including PPG/ECG acquisition, PTT estimation, Finapres calibration, Pan--Tompkins detection, and AAMI/BHS-style evaluation. \\

\midrule

\textbf{Minimal fix} 
& Remove the plate from the radar-subject line of sight, or redesign the study as a characterized through-material sensing experiment with a physically appropriate obstacle. 
& Use an ethically appropriate non-invasive reference for healthy-subject experiments, or restrict invasive validation to clinically indicated patients and revise the performance claim accordingly. \\

\bottomrule
\end{tabularx}
\caption{
\textbf{Compact annotation summary for the two false-negative cases.}
The table summarizes the gold labels, fatal triggers, missed feasibility constraints, and minimal fixes for the two representative false-negative cases.
}
\label{tab:app_false_negative_case_summary}
\end{table*}

\paragraph{Why LLMs miss the violation.}
This case is difficult because the protocol contains many surface cues associated with a legitimate biomedical signal-processing study: PPG and ECG acquisition, PTT computation, Pan--Tompkins R-peak detection, second-derivative PPG onset detection, Finapres-based calibration, and AAMI/BHS-style evaluation language.
These terms make the protocol appear methodologically rigorous.
A model can therefore accept the protocol by recognizing familiar components of cuffless blood-pressure research, while failing to audit whether the proposed validation reference is clinically and ethically feasible for the stated subject population and activity setting.
The false-negative error reflects the same pattern as Case A: the model rewards expert-like procedural scaffolding, but under-checks the feasibility condition that makes the experiment executable.

\paragraph{Minimal fix.}
A feasible version should replace routine invasive arterial-line validation in healthy volunteers with an ethically appropriate reference.
For non-clinical healthy-subject experiments, the protocol could use validated non-invasive references such as repeated cuff-based measurements or a continuous finger-cuff system, while clearly acknowledging the limitations of those references.
Alternatively, invasive arterial-line validation should be restricted to patients who already have clinically indicated arterial catheters under an approved clinical protocol.
The performance claim should also be calibrated to the validation setting, subject variability, calibration duration, and reference uncertainty, rather than specifying an unrealistically tight expected error target.

\subsection{Cross-case Summary}
\label{app:false_negative_case_summary}

\paragraph{Takeaway.}
The two cases illustrate a shared false-negative mechanism in \textsc{PhysDox}.
Both protocols are written as fluent, technically detailed experimental procedures and contain many components that resemble legitimate biomedical sensing studies.
However, the fatal violation in each case lies upstream of the model's preferred evidence surface.
In Case A, the protocol is invalid because the sensing channel is physically blocked before any radar signal-processing step can succeed.
In Case B, the protocol is invalid because the validation reference and expected performance claim are not executable for the stated healthy-subject experiment.
These failures are therefore not caused by missing terminology or incomplete descriptions.
Instead, they arise because current LLMs tend to reward complete experimental scaffolds while under-checking whether the core physical, clinical, or measurement premise is feasible.


\clearpage
\section{False-alarm Analysis}
\label{app:false_alarm}

\paragraph{Purpose.}
The main paper focuses on fatal-violation detection and false-negative cases, where a model fails to identify a protocol that contains a fatal feasibility violation.
However, a reliable protocol auditor must also avoid the opposite failure mode: over-rejecting protocols that are physically valid.
We therefore conduct an additional false-alarm analysis on the 232 ground-truth valid protocols in the Gold set.
This analysis complements the fatal-gate results by measuring whether a model can preserve feasible protocols rather than conservatively flagging them as problematic.

\paragraph{Metric.}
We define the \textit{valid false-alarm rate} as the fraction of ground-truth valid protocols that are incorrectly predicted as non-valid:

\[
\mathrm{FAR}_{\mathrm{valid}}
=
\frac{
N_{\mathrm{FA}}
}{
N_{\mathrm{valid}}
},
\]
where $N_{\mathrm{FA}}$ is the number of ground-truth valid protocols predicted as non-valid, and $N_{\mathrm{valid}}$ is the total number of ground-truth valid protocols.

Lower values indicate better preservation of feasible protocols.
For this analysis, a false alarm occurs when a ground-truth valid protocol is predicted as either \textit{minor issue} or \textit{fatal violation}.
When a model returns an empty response or an API error, we report it separately rather than counting it as a false alarm unless the output explicitly predicts a non-valid label.

\paragraph{Evaluation scope.}
This analysis is restricted to ground-truth \textit{valid} protocols and therefore measures a different failure mode from fatal-violation recall.
It should not be interpreted as a standalone measure of protocol-auditing quality.
A model can obtain a low valid false-alarm rate by being conservative and rarely flagging protocols as problematic, but such behavior may also reduce fatal-gate recall on truly invalid protocols.
Conversely, a model with high fatal recall may over-reject valid protocols if its severity boundary is poorly calibrated.
We therefore use this experiment to complement the main results: fatal-gate recall measures whether dangerous protocols are retained for diagnosis, whereas valid false-alarm rate measures whether feasible protocols are rejected.

\begin{figure*}[!t]
    \centering
    \includegraphics[width=1\linewidth]{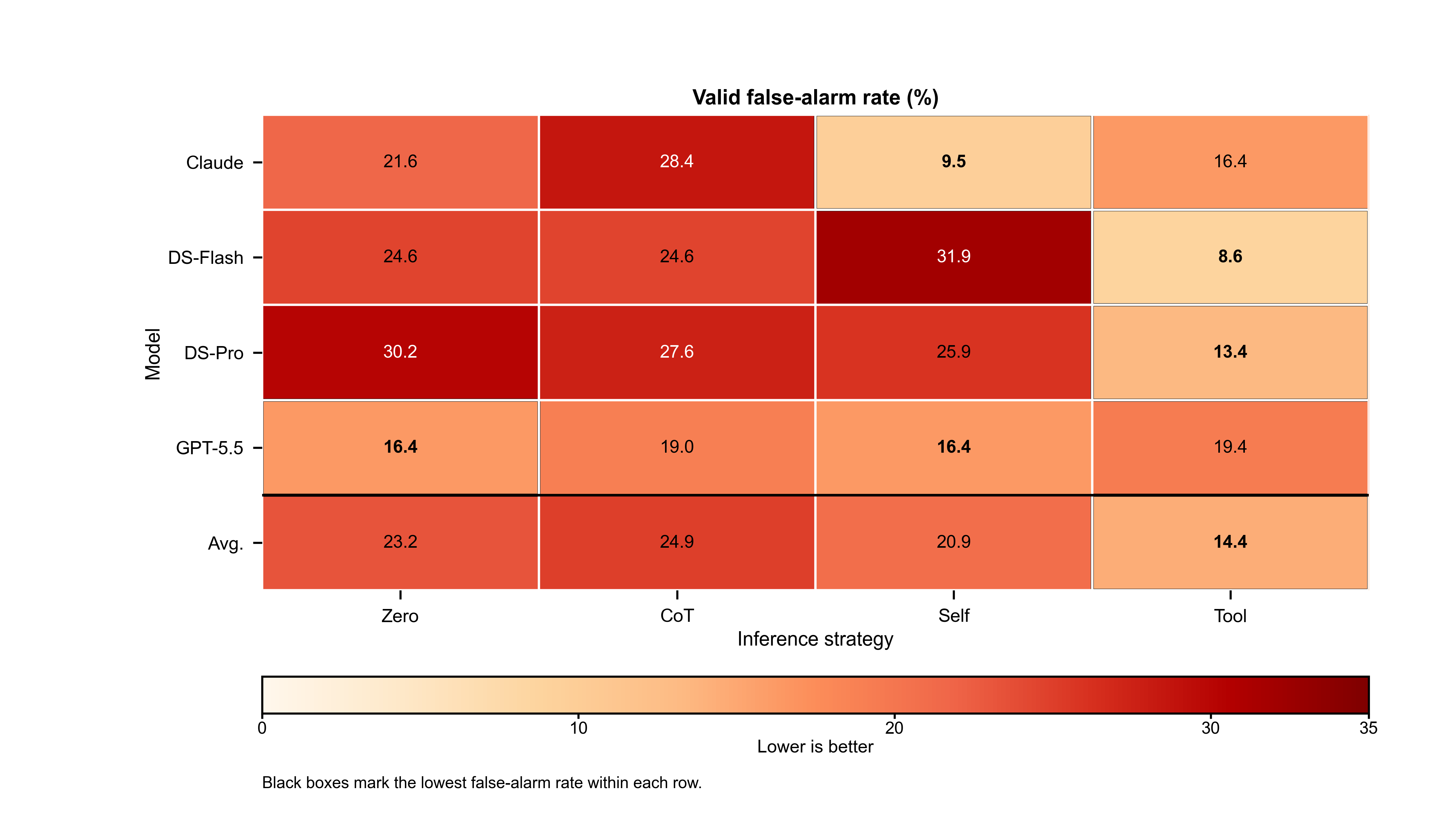}
    \caption{
    \textbf{Valid false-alarm rate across models and inference strategies.}
    Each cell reports the percentage of Gold-set valid protocols incorrectly predicted as non-valid.
    Lower values indicate better preservation of feasible protocols.
    The heatmap shows that false-alarm behavior is highly model- and strategy-dependent.
    }
    \label{fig:app_valid_false_alarm_heatmap}
\end{figure*}

\paragraph{Visualization.}
Table~\ref{tab:app_valid_false_alarm} provides the exact false-alarm counts and rates.
To make the model--strategy pattern easier to inspect, Figure~\ref{fig:app_valid_false_alarm_heatmap} visualizes the valid false-alarm rate as a heatmap.
Rows correspond to model families and columns correspond to inference strategies.
Darker cells indicate a higher tendency to over-reject valid protocols.

\paragraph{Handling empty outputs.}
For DeepSeek V4 Pro, several runs return empty responses or API-level failures.
We report these cases separately in Table~\ref{tab:app_valid_false_alarm} to avoid conflating system-level failures with semantic false alarms.
They are excluded from the false-alarm numerator unless the model explicitly predicts a non-valid label, but they are still shown because unstable outputs also affect practical deployability.

\paragraph{Results.}
Table~\ref{tab:app_valid_false_alarm} shows that valid protocols are frequently over-rejected by current LLMs.
Across the four model variants included in the false-alarm analysis (excluding the MiMo variants), the average valid false-alarm rate is 20.85\%, indicating that approximately one in five valid protocols is incorrectly flagged as problematic.
This confirms that Stage-1 severity detection is miscalibrated in both directions: models can miss fatal violations, as shown in the main paper, but they can also over-reject physically feasible protocols.
The effect of inference strategies is also non-uniform.
Tool-augmented prompting achieves the lowest average false-alarm rate, 14.44\%, suggesting that it can reduce over-conservative rejection on valid protocols.
However, this does not imply uniformly better end-to-end auditing, because tool-augmented settings can still suffer from weaker fatal-gate recall in the main experiment.
CoT prompting yields the highest average false-alarm rate, 24.90\%, suggesting that longer reasoning traces may amplify suspicion rather than improve severity calibration.
Self-consistency is model-dependent: it substantially reduces false alarms for Claude Opus 4.7, but increases false alarms for DeepSeek V4 Flash.
These results show that common inference-time strategies do not provide a reliable solution to the severity-calibration problem.

\paragraph{Takeaway.}
Complementing the main paper's false-negative study, this analysis highlights a critical LLM failure: the inability to balance detecting fatal violations against preserving valid protocols. Current LLMs act as miscalibrated severity classifiers, struggling to differentiate valid, minor, and fatally flawed protocols. This reinforces \textsc{PhysDox}'s core finding: feasibility assessment demands holistic, calibrated reasoning, not superficial keyword matching or verbose rationales.

\begin{table*}[!t]
\centering
\small
\setlength{\tabcolsep}{5pt}
\renewcommand{\arraystretch}{1.12}
\begin{tabular}{llrrrrr}
\toprule
\textbf{Model} 
& \textbf{Strategy} 
& \textbf{$N_{\mathrm{valid}}$} 
& \textbf{Correct Valid} 
& \textbf{False Alarms} 
& \textbf{Empty / API Errors} 
& \textbf{False-alarm Rate (\%)} \\
\midrule

\multirow{4}{*}{Claude Opus 4.7}
& Zero-shot & 232 & 182 & 50 & 0 & 21.55 \\
& CoT & 232 & 166 & 66 & 0 & 28.45 \\
& Self-Cons. & 232 & 206 & 22 & 0 & \textbf{9.48} \\
& Tool-Aug. & 232 & 194 & 38 & 0 & 16.38 \\

\midrule
\multirow{4}{*}{DeepSeek V4 Flash}
& Zero-shot & 232 & 175 & 57 & 0 & 24.57 \\
& CoT & 232 & 175 & 57 & 0 & 24.57 \\
& Self-Cons. & 232 & 158 & 74 & 0 & 31.90 \\
& Tool-Aug. & 232 & 212 & 20 & 0 & \textbf{8.62} \\

\midrule
\multirow{4}{*}{DeepSeek V4 Pro}
& Zero-shot & 232 & 139 & 70 & 23 & 30.17 \\
& CoT & 232 & 116 & 64 & 52 & 27.59 \\
& Self-Cons. & 232 & 172 & 60 & 0 & 25.86 \\
& Tool-Aug. & 232 & 180 & 31 & 21 & \textbf{13.36} \\

\midrule
\multirow{4}{*}{GPT-5.5}
& Zero-shot & 232 & 194 & 38 & 0 & \textbf{16.38} \\
& CoT & 232 & 188 & 44 & 0 & 18.97 \\
& Self-Cons. & 232 & 194 & 38 & 0 & \textbf{16.38} \\
& Tool-Aug. & 232 & 187 & 45 & 0 & 19.40 \\

\midrule
\multirow{4}{*}{\textbf{Strategy Avg.}}
& Zero-shot & -- & -- & -- & -- & 23.17 \\
& CoT & -- & -- & -- & -- & 24.90 \\
& Self-Cons. & -- & -- & -- & -- & 20.90 \\
& Tool-Aug. & -- & -- & -- & -- & \textbf{14.44} \\

\midrule
\textbf{Overall Avg.}
& All settings & -- & -- & -- & -- & \textbf{20.85} \\

\bottomrule
\end{tabular}
\caption{
\textbf{False-alarm analysis on Gold-set valid protocols.}
The table reports model behavior on the 232 ground-truth valid protocols in the Gold set.
A false alarm occurs when a valid protocol is predicted as non-valid.
The valid false-alarm rate is computed over all 232 valid protocols.
Empty outputs and API errors are reported separately and are not counted as false alarms unless the model explicitly predicts a non-valid label.
Bold values indicate the lowest false-alarm rate within each model block.
}
\label{tab:app_valid_false_alarm}
\end{table*}


\begin{table*}[t]
\centering
\tiny
\setlength{\tabcolsep}{3.2pt}
\renewcommand{\arraystretch}{1.08}
\resizebox{\textwidth}{!}{%
\begin{tabular}{llr cccc cccc cccc}
\toprule
\multirow{2}{*}{\textbf{Model}}
& \multirow{2}{*}{\textbf{Strategy}}
& \multirow{2}{*}{\textbf{Gate N}}
& \multicolumn{4}{c}{\textbf{Oracle}}
& \multicolumn{4}{c}{\textbf{Gated}}
& \multicolumn{4}{c}{\textbf{End-to-End}} \\
\cmidrule(lr){4-7}
\cmidrule(lr){8-11}
\cmidrule(lr){12-15}
& & 
& \textbf{Acc.} & \textbf{BAcc.} & \textbf{M-F1} & \textbf{Top-2}
& \textbf{Acc.} & \textbf{BAcc.} & \textbf{M-F1} & \textbf{Top-2}
& \textbf{Acc.} & \textbf{BAcc.} & \textbf{M-F1} & \textbf{Top-2} \\
\midrule

GPT-5.5 & Zero-Shot & 313 & 50.7 & 55.0 & 49.8 & 85.9 & 52.1 & 56.2 & 51.6 & 87.2 & 43.2 & 48.1 & 46.6 & 72.4 \\
GPT-5.5 & CoT & 293 & 52.3 & 55.9 & 51.0 & 84.4 & 53.2 & 55.9 & 52.3 & 84.6 & 41.4 & 44.6 & 45.9 & 65.8 \\
GPT-5.5 & Self-Cons. & 316 & 53.1 & 57.4 & 52.8 & 84.1 & 54.1 & 58.4 & 54.5 & 85.4 & 45.4 & 50.0 & 49.5 & 71.6 \\
GPT-5.5 & Tool-Aug. & 282 & 56.0 & 58.9 & 55.5 & 86.7 & 58.2 & 61.4 & 58.6 & 88.7 & 43.5 & 46.5 & 50.1 & 66.3 \\
\midrule

Claude Opus 4.7 & Zero-Shot & 245 & 58.1 & 59.3 & 56.4 & 91.0 & 65.3 & 67.9 & 64.2 & 93.9 & 42.4 & 44.4 & 51.3 & 61.0 \\
Claude Opus 4.7 & CoT & 217 & 55.7 & 55.2 & 53.2 & 88.3 & 65.0 & 65.8 & 63.7 & 90.8 & 37.4 & 39.9 & 48.3 & 52.3 \\
Claude Opus 4.7 & Self-Cons. & 255 & 58.9 & 59.0 & 56.6 & 90.7 & 65.1 & 67.0 & 63.9 & 95.3 & 44.0 & 44.9 & 51.8 & 64.5 \\
Claude Opus 4.7 & Tool-Aug. & 229 & 63.9 & 60.5 & 60.8 & 92.0 & 68.1 & 67.1 & 66.9 & 94.3 & 41.4 & 42.2 & 51.6 & 57.3 \\
\midrule

DS V4 Flash & Zero-Shot & 283 & 39.0 & 44.5 & 36.8 & 82.0 & 39.9 & 45.5 & 38.2 & 82.3 & 30.0 & 34.2 & 32.3 & 61.8 \\
DS V4 Flash & CoT & 277 & 40.3 & 46.9 & 38.6 & 81.7 & 43.3 & 49.7 & 42.6 & 82.7 & 31.8 & 37.0 & 36.1 & 60.7 \\
DS V4 Flash & Self-Cons. & 301 & 38.7 & 44.5 & 37.4 & 83.8 & 41.9 & 46.7 & 40.2 & 86.7 & 33.4 & 38.5 & 35.5 & 69.2 \\
DS V4 Flash & Tool-Aug. & 245 & 43.5 & 43.7 & 40.5 & 84.6 & 43.3 & 47.2 & 42.5 & 87.8 & 28.1 & 30.1 & 33.2 & 57.0 \\
\midrule

DS V4 Pro & Zero-Shot & 215 & 48.0 & 54.8 & 48.3 & 84.9 & 50.7 & 56.7 & 51.2 & 87.0 & 28.9 & 33.8 & 37.8 & 49.6 \\
DS V4 Pro & CoT & 223 & 45.9 & 50.7 & 46.2 & 84.1 & 49.8 & 53.6 & 50.6 & 87.0 & 29.4 & 33.1 & 37.5 & 51.5 \\
DS V4 Pro & Self-Cons. & 236 & 46.4 & 53.9 & 48.6 & 87.3 & 49.6 & 57.1 & 52.2 & 90.7 & 31.0 & 36.9 & 40.4 & 56.8 \\
DS V4 Pro & Tool-Aug. & 169 & 48.0 & 51.9 & 50.2 & 88.3 & 53.3 & 60.2 & 55.7 & 88.8 & 23.9 & 27.5 & 34.9 & 39.8 \\
\midrule

MiMO v2.5 Pro & Zero-Shot & 142 & 52.0 & 52.2 & 49.8 & 86.7 & 61.3 & 62.9 & 62.5 & 90.1 & 23.1 & 26.3 & 35.8 & 34.0 \\
MiMO v2.5 Pro & CoT & 152 & 54.9 & 56.8 & 53.9 & 86.7 & 61.8 & 64.4 & 61.4 & 87.5 & 24.9 & 28.2 & 37.3 & 35.3 \\
MiMO v2.5 Pro & Self-Cons. & 137 & 57.0 & 58.6 & 55.9 & 89.9 & 67.2 & 70.3 & 69.3 & 95.6 & 24.4 & 28.3 & 38.7 & 34.7 \\
MiMO v2.5 Pro & Tool-Aug. & 134 & 55.4 & 53.8 & 53.5 & 89.1 & 61.9 & 63.7 & 64.3 & 94.0 & 22.0 & 24.6 & 34.6 & 33.4 \\
\midrule

MiMO v2.5 & Zero-Shot & 144 & 49.9 & 55.2 & 49.9 & 86.7 & 56.9 & 62.8 & 57.8 & 88.2 & 21.8 & 26.5 & 34.1 & 33.7 \\
MiMO v2.5 & CoT & 163 & 50.1 & 53.8 & 49.3 & 84.1 & 57.7 & 63.1 & 58.1 & 87.1 & 24.9 & 29.1 & 36.9 & 37.7 \\
MiMO v2.5 & Self-Cons. & 163 & 52.3 & 57.1 & 52.0 & 86.2 & 63.2 & 68.3 & 64.6 & 90.2 & 27.3 & 32.5 & 41.1 & 39.0 \\
MiMO v2.5 & Tool-Aug. & 154 & 54.4 & 51.6 & 51.5 & 87.8 & 59.7 & 60.2 & 60.3 & 89.6 & 24.4 & 25.1 & 35.2 & 36.6 \\

\bottomrule
\end{tabular}%
}
\caption{
\textbf{Full Stage-2 three-way diagnostic results on the Gold set.}
Oracle evaluates diagnosis on all ground-truth fatal samples.
Gated evaluates diagnosis only on fatal samples retained by Stage-1 gating.
End-to-End evaluates the full two-stage pipeline.
All metrics except Gate N are percentages.
}
\label{tab:app_gold683_three_full}
\end{table*}

\section{Full Stage-2 Diagnostic Results}
\label{app:full_stage2_results}

This appendix provides the full Stage-2 three-way diagnostic results on the Gold set.
The main paper reports a subset of two-stage metrics to highlight the severity-gate bottleneck, while Table~\ref{tab:app_gold683_three_full} expands the Stage-2 diagnosis results with accuracy, balanced accuracy, macro-F1, and Top-2 accuracy under Oracle, Gated, and End-to-End settings.

\paragraph{Reading the table.}
Oracle evaluates diagnosis on all ground-truth fatal-violation protocols, isolating whether a model can identify the coarse physical-error category once the protocol is already known to be fatal.
Gated evaluates diagnosis only on the ground-truth fatal protocols retained by Stage-1 fatal gating; Gate N reports the retained count out of 377 fatal protocols.
End-to-End evaluates the full two-stage pipeline over all ground-truth fatal protocols and therefore penalizes both missed fatal cases and incorrect downstream diagnosis.
Thus, Gated scores should be interpreted as conditional diagnostic performance, whereas End-to-End scores reflect practical protocol-auditing performance.

\paragraph{Observation.}
Table~\ref{tab:app_gold683_three_full} confirms that strong conditional diagnosis does not necessarily translate into reliable end-to-end auditing. Several settings achieve high Gated accuracy or Top-2 accuracy after Stage-1 filtering, but their End-to-End performance remains substantially lower because many fatal protocols are not routed to Stage 2. These results reinforce the main-paper finding that severity-gate calibration, rather than downstream taxonomy classification alone, is the key bottleneck.

This systemic bottleneck is vividly demonstrated by Claude Opus 4.7 (Tool-Augmented), which achieves the highest Gated and Fine-12 oracle metrics yet fails to maintain this advantage end-to-end due to anaemic fatal-gate recall. 
Consequently, these results provide empirical validation for our two-stage task formulation, confirming that resolving the upstream severity calibration deadlock is far more critical for scientific safety than merely expanding downstream diagnostic vocabularies.



\begin{table*}[t]
\centering
\small
\begin{tabular}{llll}
\toprule
Model & Family / regime & Size metadata & Role in evaluation \\
\midrule
GPT-5.5 & GPT-family hosted model & Undisclosed & Frontier general-purpose LLM \\
Claude Opus 4.7 & Claude-family hosted model & Undisclosed & Strong reasoning-oriented LLM \\
DeepSeek V4 Flash & DeepSeek-family efficient variant & Undisclosed & Cost-efficient deployment-oriented model \\
DeepSeek V4 Pro & DeepSeek-family stronger variant & Undisclosed & Stronger same-family comparison point \\
MiMo v2.5 Pro & MiMo-family stronger variant & Undisclosed & Strong MiMo-family comparison point \\
MiMo v2.5 & MiMo-family lightweight variant & Undisclosed & Lightweight MiMo-family comparison point \\
\bottomrule
\end{tabular}
\caption{Model coverage in \textsc{PhysDox}. The table documents the role of each evaluated model rather than presenting a parameter-scaling study. Exact parameter counts and training details are marked as undisclosed when they are not publicly available or not directly verifiable from our evaluation setup.}
\label{tab:model_coverage}
\end{table*}

\section{Model Selection, Scaling Boundaries, and Silver Provenance}
\label{app:model_selection}

\paragraph{Selection rationale.} Our goal is not to benchmark every available LLM architecture, but to test whether physical-feasibility auditing failures persist across representative current model families and inference-time strategies. We therefore select six LLM variants spanning multiple provider families, access regimes, and capability levels. The evaluation includes stronger general-purpose models, lightweight or cost-efficient variants, and models evaluated under the same four inference strategies: zero-shot, chain-of-thought prompting, self-consistency, and tool-augmented prompting.

\paragraph{Model metadata and evaluation roles.}
Table~\ref{tab:model_coverage} summarizes the model coverage used in \textsc{PhysDox}. We report only metadata that is available or directly relevant to our evaluation. For models whose exact parameter counts or training details are not publicly disclosed, we mark these fields as undisclosed rather than estimating them. We also distinguish model access from weight inspection: in this study, models are evaluated through their available inference endpoints, and we do not rely on local checkpoint inspection or parameter-level analysis.

\paragraph{Why we do not claim a scaling law.} A controlled scaling-law analysis would require a matched family of models with known parameter counts, comparable training data, identical deployment constraints, and sufficiently dense size coverage. The current evaluation does not satisfy these requirements because several API models do not disclose model size, and the evaluated systems differ in training data, alignment procedure, inference stack, and tool-use behavior. We therefore avoid claiming that physical-feasibility auditing is or is not an emergent capability of scale. Instead, our results show that severity-calibration failure appears across the evaluated model families and is not resolved by common inference-time strategies.

\paragraph{Architecture and domain-specialized models.}
Our benchmark does not cover every architecture lineage, such as Llama-family open-weight models, nor models specifically tuned on code, mathematics, physics, or scientific corpora. This is a limitation of the present study. Such models would be informative for testing whether explicit scientific or numerical training improves physical feasibility auditing. We leave this broader architecture and specialization study to future benchmark expansions. The present conclusions should therefore be interpreted as evidence of a persistent failure pattern across the evaluated systems, not as an exhaustive claim over all possible LLM architectures.

\paragraph{Interpretation boundaries.}
The evaluated model suite supports three conclusions. First, physical-feasibility auditing remains difficult for several strong contemporary LLM families under a shared evaluation protocol. Second, the main bottleneck is severity calibration rather than only downstream fatal-error taxonomy classification. Third, common inference-time strategies, including longer rationales, self-consistency, and generic tool augmentation, do not reliably resolve this bottleneck. The evaluation does not support stronger claims about all possible open-weight architectures, all domain-specialized scientific models, or the scaling behavior of physical auditing ability with parameter count.

\paragraph{Relation to the Silver set.} The Silver set is used for large-scale trend analysis rather than as the basis for safety conclusions. This distinction is important because the Silver release includes targeted MiMo-generated candidates used to fill label and fatal-constraint quotas. All main safety claims are therefore based on the expert-curated Gold set, which is disjoint from the Silver set and is used as the primary benchmark for model comparison.


\newpage
\section{Silver 5k Data Curation Pipeline}
\label{app:silver_pipeline}

\paragraph{Purpose.}
This appendix documents the construction of the Silver 5k release.
The goal is to preserve only fully specified, unique, hardware-grounded protocols while using targeted generation to fill balanced label and fatal-constraint quotas.
The Silver set is therefore not a raw model-generated pool, but a filtered and frozen benchmark split designed for large-scale trend evaluation.

\subsection{Construction Overview}
\label{app:silver_pipeline_overview}

\begin{figure}[t]
    \centering
    \includegraphics[width=1\linewidth]{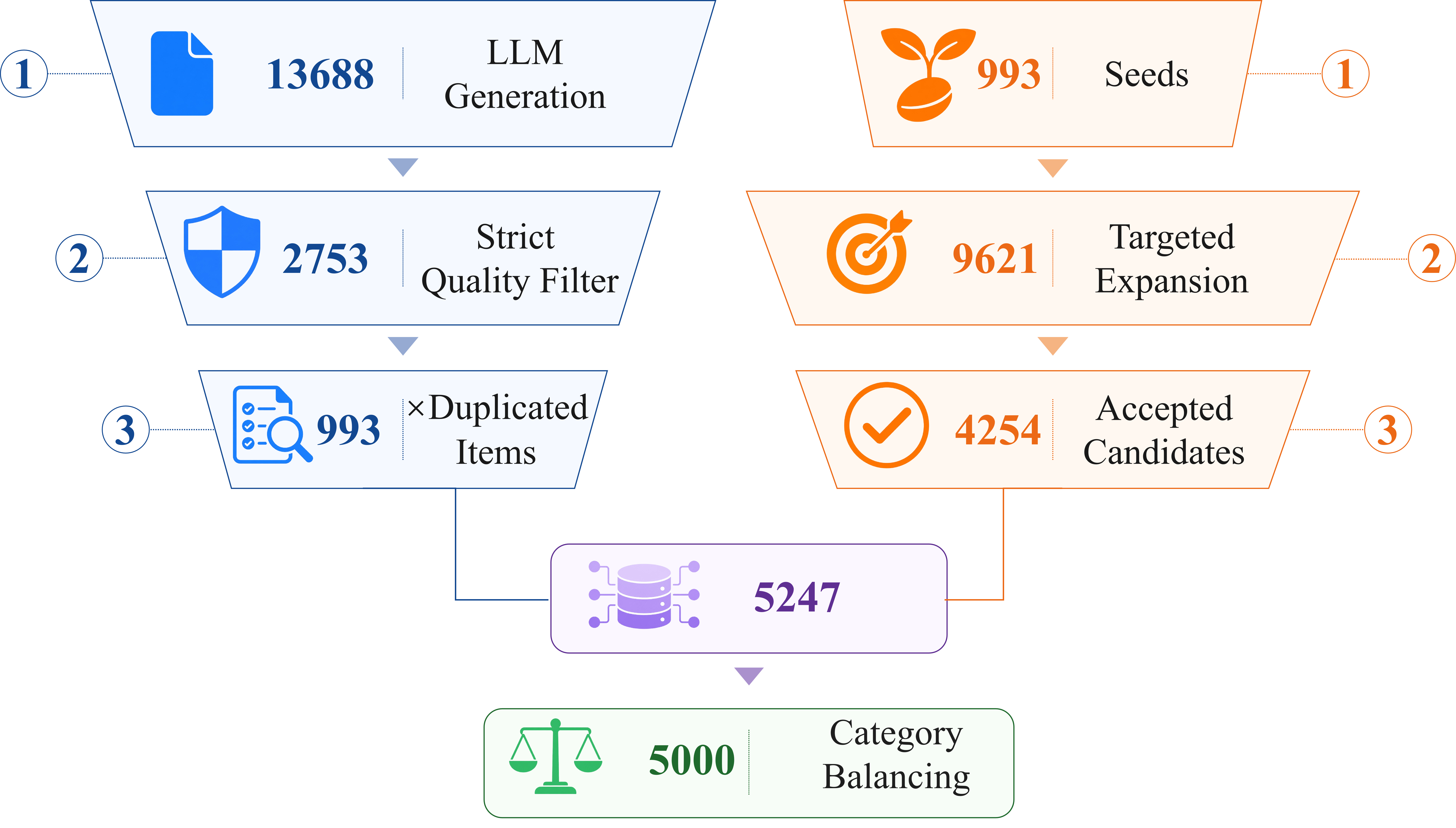}
    \caption{
\textbf{Silver 5k data curation pipeline.}
The pipeline starts from 13,688 candidate protocols, applies strict quality filtering and deduplication, supplements missing quotas through targeted MiMO generation, and freezes the final 5,000-sample Silver release through deterministic quota sampling.
}
    \label{figR5}
\end{figure}

Figure~\ref{figR5} summarizes the Silver-set curation pipeline.
We start from 13,688 candidate protocols aggregated from existing Silver candidates and historical generated pools, which include valid, minor-issue, and fatal-violation candidates before strict public-release filtering.
We first apply quality and leakage filters to remove protocols with missing mandatory sections, insufficient section content, wrapper text, prompt leakage, placeholder artifacts, or incomplete label metadata.
This step retains 2,753 candidates.
We then normalize whitespace and remove duplicate protocol texts, reducing the quality-pass pool to 993 unique clean seeds.
Because the filtered seed pool is insufficient to satisfy the planned label and fatal-constraint quotas, we perform 9,621 targeted MiMO v2.5 Pro generation attempts to supplement underrepresented labels and constraint categories.
This process yields 4,254 accepted generated candidates.
The 993 clean seeds and 4,254 accepted generated candidates are merged into a strict candidate pool of 5,247 protocols.
Finally, we perform deterministic quota sampling with seed 42 to freeze the final 5,000-sample Silver release.

\begin{table*}[t]
\centering
\small
\setlength{\tabcolsep}{5pt}
\renewcommand{\arraystretch}{1.10}

\begin{tabularx}{\textwidth}{
    @{}
    >{\raggedright\arraybackslash}p{0.18\textwidth}
    >{\raggedleft\arraybackslash}p{0.07\textwidth}
    >{\raggedright\arraybackslash}X
    >{\raggedright\arraybackslash}p{0.25\textwidth}
    @{}
}
\toprule
\textbf{Stage} & \textbf{Count} & \textbf{Operation} & \textbf{Notes} \\
\midrule

Initial candidate pool
& 13,688
& Aggregate existing Silver candidates and historical generated pools
& Includes valid, minor, and fatal candidates before strict public-release filtering. \\

Quality-pass raw candidates
& 2,753
& Apply protocol quality and leakage filters
& 10,935 candidates rejected by the first triggered quality rule. \\

Unique clean seed pool
& 993
& Normalize whitespace and remove duplicate protocol text
& 1,760 duplicate protocol rows removed after quality filtering. \\

Targeted MiMo generation attempts
& 9,621
& Generate valid, minor, and fatal candidates to fill missing label and constraint quotas
& MiMo v2.5 Pro with batched targeted generation. \\

Accepted generated candidates
& 4,254
& Keep generated candidates passing the same quality and uniqueness filters
& Computed as 9,621 attempts minus 5,367 rejected attempts. \\

Strict pool before final sampling
& 5,247
& Combine 993 unique clean seeds with 4,254 accepted generated candidates
& Used as the strict unique pool for final quota sampling. \\

Frozen Silver 5k release
& 5,000
& Deterministic quota sampling with seed 42
& Frozen at 2,250 fatal, 1,500 valid, and 1,250 minor samples. \\

\bottomrule
\end{tabularx}

\caption{
\textbf{Detailed construction funnel for the Silver 5k release.}
The table reports each major curation stage, the number of candidates retained,
and the operation applied at that stage.
}
\label{tab:app_silver_funnel}
\end{table*}
\subsection{Quality Filtering and Rejection Analysis}
\label{app:silver_quality_filtering}

Quality filtering is deliberately strict.
Candidates are rejected if they lack mandatory sections, contain wrappers or prompt leakage, are too short, contain weak section content, duplicate another protocol after normalization, or violate label-specific metadata rules.
For generated candidates, the same filtering and uniqueness checks are applied before they can enter the strict candidate pool.
Table~\ref{tab:app_silver_rejection} summarizes the dominant rejection reasons.

\begin{table}[t]
\centering
\small
\setlength{\tabcolsep}{4pt}
\renewcommand{\arraystretch}{1.08}
\begin{tabular}{lr}
\toprule
\textbf{Rejection reason} & \textbf{Count} \\
\midrule
\multicolumn{2}{l}{\textit{Source candidates}} \\
Missing mandatory section headers & 5,199 \\
Forbidden wrapper / prompt leakage & 3,065 \\
Protocol length below 180 words & 753 \\
Weak data acquisition section & 663 \\
Weak objective section & 658 \\
Other quality or metadata failures & 597 \\
\midrule
\multicolumn{2}{l}{\textit{Generated candidates}} \\
JSON parse failure & 3,381 \\
Protocol length below 180 words & 1,832 \\
Weak data acquisition section & 82 \\
Forbidden wrapper / prompt leakage & 32 \\
Other section-quality failures & 40 \\
\bottomrule
\end{tabular}
\caption{
\textbf{Main rejection reasons during Silver-set curation.}
}
\label{tab:app_silver_rejection}
\end{table}

\subsection{Final Silver-set Composition}
\label{app:silver_composition}

The final Silver release contains 5,000 unique protocols with fixed label quotas and broad coverage across six biomedical sensing domains.
As shown in Table~\ref{tab:app_silver_composition}, the release contains 2,250 fatal violations, 1,500 valid protocols, and 1,250 minor-issue protocols.
Fatal-violation samples are approximately balanced across the 12 fatal constraint keys, with each key contributing 187--188 samples.

\begin{table}[t]
\centering
\small
\setlength{\tabcolsep}{5pt}
\renewcommand{\arraystretch}{1.08}
\begin{tabular}{lrr}
\toprule
\textbf{Category} & \textbf{Count} & \textbf{Share} \\
\midrule
\multicolumn{3}{l}{\textit{Severity labels}} \\
Fatal violation & 2,250 & 45.0\% \\
Valid & 1,500 & 30.0\% \\
Minor issue & 1,250 & 25.0\% \\
\midrule
\multicolumn{3}{l}{\textit{Domains}} \\
IMU motion & 1,019 & 20.4\% \\
sEMG gesture & 1,009 & 20.2\% \\
EEG BCI & 827 & 16.5\% \\
PPG blood pressure & 732 & 14.6\% \\
Radar vital signs & 728 & 14.6\% \\
Throat-microphone speech & 685 & 13.7\% \\
\bottomrule
\end{tabular}
\caption{
\textbf{Final composition of the frozen Silver 5k release.}
The release contains 5,000 unique protocols with fixed label quotas and broad domain coverage across six biomedical sensing settings.
}
\label{tab:app_silver_composition}
\end{table}

\subsection{Cleaning and Generation Rules}
\label{app:silver_cleaning_rules}

\paragraph{Protocol format validation.}
A protocol can enter the strict unique pool only if it satisfies release-level quality checks.
Each protocol must contain exactly seven mandatory sections: \textit{Protocol}, \textit{Objective}, \textit{Equipment}, \textit{Sensor Placement \& Configuration}, \textit{Data Acquisition Procedure}, \textit{Signal Processing \& Validation}, and \textit{Safety \& Feasibility}.
The protocol must contain at least 180 words, and each section must contain substantive content rather than placeholder text.
Protocol text must also be unique after whitespace normalization.

\paragraph{Leakage and wrapper filtering.}
We reject candidates containing wrappers, placeholders, prompt leakage, instruction traces, or benchmark-construction language.
Examples include bracket placeholders, ``source protocol text'', ``procedure below'', ``as written above'', ``original protocol'', ``rewritten protocol'', ``I will rewrite'', ``output only'', ``required headers'', and ``benchmark construction''.
This filtering is designed to remove model-generation artifacts and ensure that the released protocols resemble standalone experimental methods rather than prompt-completion traces.

\paragraph{Label-specific metadata validation.}
Label-specific metadata is checked before final release.
Valid protocols must not retain violation metadata.
Minor-issue samples must include complete metadata fields describing the violated best-practice principle, why the protocol remains physically feasible, the minimal fix, and what changed.
Fatal-violation samples must be tied to one of the 12 fatal constraint keys.
The final release is validated to contain 5,000 unique IDs, 5,000 unique protocol texts, exact label quotas, exact fatal-constraint quotas, no remaining quality issues, no wrapper hits, no bracket-placeholder hits, and no protocols under the minimum-length threshold.

\paragraph{Targeted MiMo generation.}
Targeted MiMo generation is used only after filtering and deduplication, and only to fill missing label or fatal-constraint quotas.
Valid, minor-issue, and fatal-violation templates share the same seven-section output format.
For fatal-violation, each call is conditioned on a domain, sensors, target signals, typical parameters, a study variant, and one fatal constraint key.
The generation prompt requires the protocol to remain fluent and plausible as an experimental methods section while containing exactly one fatal flaw tied to the specified constraint.
Generated candidates are not automatically admitted into the benchmark; they must pass the same quality, leakage, metadata, and uniqueness checks as source candidates.

\paragraph{Takeaway.}
Designed as a controlled evaluation split rather than a raw corpus, the Silver 5k set undergoes rigorous curation. The pipeline filters invalid, duplicated, or leaked protocols, enforces structural consistency, and employs targeted MiMO generation to fulfill specific quotas. Deterministic sampling ensures balanced severity and constraint distributions. This split facilitates large-scale trend analysis, complementing the expert-curated Gold benchmark used for definitive model comparison.

\end{document}